\documentclass[sn-mathphys,Numbered]{sn-jnl}% Math and Physical Sciences Reference Style
\usepackage{graphicx}%
\usepackage{multirow}%
\usepackage{amsmath,amssymb,amsfonts}%
\usepackage{amsthm}%
\usepackage{mathrsfs}%
\usepackage[title]{appendix}%
\usepackage{xcolor}%
\usepackage{textcomp}%
\usepackage{manyfoot}%
\usepackage{booktabs}%
\usepackage{algorithm}%
\usepackage{algorithmicx}%
\usepackage{algpseudocode}%
\usepackage{listings}%
\usepackage{siunitx}
\usepackage{lineno}
\usepackage{ulem}
\usepackage{url}
\usepackage{longtable}
\usepackage{multirow}
\usepackage{caption}
\usepackage{tabularx}
\usepackage{pdflscape}
\usepackage{lineno}
\begin{document}

\title{Optimizing VGOS observations using an SNR-based scheduling approach}

\author*[1]{\fnm{Matthias} \sur{Schartner}}\email{mschartner@ethz.ch}

\author[2]{\fnm{Bill} \sur{Petrachenko}}\email{wtpetra@gmail.com}
\equalcont{These authors contributed equally to this work.}

\author[3]{\fnm{Mike} \sur{Titus}}\email{matitus@mit.edu}
\equalcont{These authors contributed equally to this work.}

\author[4]{\fnm{Hana} \sur{Kr\'asn\'a}}\email{hana.krasna@geo.tuwien.ac.at}
\equalcont{These authors contributed equally to this work.}

\author[3]{\fnm{John} \sur{Barrett}}\email{barrettj@mit.edu}
\equalcont{These authors contributed equally to this work.}

\author[3]{\fnm{Dan} \sur{Hoak}}\email{dhoak@mit.edu}
\equalcont{These authors contributed equally to this work.}

\author[3]{\fnm{Dhiman} \sur{Mondal}}\email{dmondal@mit.edu}
\equalcont{These authors contributed equally to this work.}

\author[5]{\fnm{Minghui} \sur{Xu}}\email{minghui.xu@gfz-potsdam.de}
\equalcont{These authors contributed equally to this work.}

\author[1]{\fnm{Benedikt} \sur{Soja}}\email{soja@ethz.ch}
\equalcont{These authors contributed equally to this work.}

\affil*[1]{\orgdiv{Institute of Geodesy and Photogrammetry}, \orgname{ETH Zurich}, \orgaddress{\street{Robert-Gnehm-Weg 15}, \city{Zürich}, \postcode{8093}, \country{Switzerland}}}

\affil[2]{\orgname{Natural Resources Canada (retired)}}

\affil[3]{\orgname{MIT Haystack Observatory}}

\affil[4]{\orgname{TU Wien}}

\affil[5]{\orgname{DeutschesGeoForschungsZentrum (GFZ) Potsdam}}


\abstract{
The geodetic and astrometric Very Long Baseline Interferometry (VLBI) community is in the process of upgrading its existing infrastructure with the VLBI Global Observing System (VGOS). 
The primary objective of VGOS is to substantially boost the number of scans per hour for enhanced parameter estimation. 
However, the current observing strategy results in fewer scans than anticipated.

During 2022, six 24-hour VGOS Research and Development (R\&D) sessions were conducted to demonstrate a proof-of-concept aimed at addressing this shortcoming. 
The new observation strategy centers around a signal-to-noise (SNR)--based scheduling approach combined with eliminating existing overhead times in existing VGOS sessions.
Two SNR-based scheduling approaches were tested during these sessions: one utilizing inter-/extrapolation of existing S/X source flux density models and another based on a newly derived source flux density catalog at VGOS frequencies. 
Both approaches proved effective, leading to a 2.3-fold increase in the number of scheduled scans per station and a 2.6-fold increase in the number of observations per station, while maintaining a high observation success rate of approximately \SIrange{90}{95}{\%}. 
Consequently, both strategies succeeded in the main objective of these sessions by successfully increasing the number of scans per hour.
The strategies described in this work can be easily applied to operational VGOS observations.

Besides outlining and discussing the observation strategy, we further provide insight into the resulting signal-to-noise ratios, and discuss the impact on the precision of the estimated geodetic parameters.
Monte Carlo simulations predicted a roughly \SI{50}{\%} increase in geodetic precision compared to operational VGOS sessions. 
The analysis confirmed that the formal errors in estimated station coordinates were reduced by \SIrange{40}{50}{\%}. 
Additionally, Earth orientation parameters showed significant improvement, with a \SIrange{40}{50}{\%} reduction in formal errors. 
}


\maketitle

\section{Introduction}\label{sec:introduction}
Very Long Baseline Interferometry (VLBI) is a cutting-edge technique in space geodesy. 
Through the synchronized observations of multiple radio telescopes strategically positioned worldwide, VLBI attains unrivaled accuracy in determining the rotation angle of the Earth about its axis and in monitoring minute variations of the orientation of the Earth's rotation vector in space. 
It furthermore contributes to the establishment of the International Terrestrial Reference Frame \citep[ITRF;][]{Altamimi2023} while also defining the International Celestial Reference Frame \citep[ICRF; ][]{Charlot2020} in its current realization.

Geodetic VLBI observations are conducted in sessions that are organized and supervised by the International VLBI Service for Geodesy and Astrometry \citep[IVS;][]{Nothnagel2017}. 
Among others, the IVS organizes the IVS-R1 and IVS-R4 series, which are global 24-hour VLBI sessions with a rapid turnaround time \citep{Thomas2024}. 
These sessions are observed using a dual-frequency mode at S/X-band (approximately \SI{2.3}{GHz} and \SI{8.6}{GHz}) and a recording rate of \SIrange{256}{512}{Mbps}. 
This way, around 20 scans per hour can be achieved. 
However, the S/X network suffers from an aging infrastructure and a severe inhomogeneity w.r.t. telescope properties, making it hard to optimize its observations. 
Since it was foreseen that the S/X infrastructure could not meet the demanding requirements posed by the Global Geodetic Observing System \citep[GGOS; ][]{Plag2009, Beutler2012}, members of the IVS decided to upgrade the S/X infrastructure.

During the design process of the successor infrastructure, tropospheric turbulences have been identified as the primary error source in geodetic VLBI \citep{Pany2011}. 
As a countermeasure, increased sampling of the troposphere at different azimuth and elevation angles over short periods is required. 
Simulations revealed that a source switching interval of approximately 30 seconds, equivalent to 120 scans per hour, will be required to reach the GGOS requirements \cite{Petrachenko2008}.
Consequently, a new telescope network utilizing fast slewing telescopes, the so-called VLBI Global Observing System (VGOS), was designed and is currently being built \citep{Niell2006, Petrachenko2009}. 
To achieve high slewing rates, a sacrifice w.r.t. the antenna diameter has to be made which impacts their sensitivity. 
To compensate for this, a new observing mode was developed utilizing four bands and an increased recording rate. 
Currently, the VGOS-mode observes four \SI{512}{MHz} wide bands A--D with center frequencies approximately at \SI{3.2}{GHz}, \SI{5.5}{GHz}, \SI{6.6}{GHz}, and \SI{10.4}{GHz} \citep{Niell2018}. 

Since 2020, the VGOS network has been operationally observing VLBI sessions within the so-called VGOS-OPS observing program. 
Telescopes participating in VGOS-OPS observe around 40 scans per hour, a factor of two better than S/X but far from the originally anticipated 120 scans per hour. 
The discrepancy can be explained by the following:
First, all scans are observed for 30 seconds straight, independent of telescope sensitivity and source brightness, while the original VGOS design document anticipated an observing time of as short as \SI{5}{s} \cite{Petrachenko2009}. 
Besides, other technological limitations exist that introduce additional overhead times. 
Together, these effects accumulate to a theoretical minimal source switching interval of \SI{68}{s}, significantly higher than the anticipated \SI{30}{s}. 

To address this shortcoming, the IVS has provided resources to observe six dedicated research and development sessions in 2022, the VGOS-R\&D program.
The VGOS-R\&D program aimed to develop appropriate methodologies and concepts to increase the number of scans per hour. 
Most importantly, the feasibility of an optimized, SNR-based observation strategy featuring shorter observation times was explored, together with a careful evaluation and elimination of existing technological limitations as a secondary measure. 

In this work, we will report on the actions and investigations of the VGOS-R\&D sessions. 
Section~\ref{sec:data} describes the VGOS sessions, Session~\ref{sec:method} describes the methodologies, Session~\ref{sec:results} discusses the evaluation metrics and presents results, Section~\ref{sec:conclusion} concludes the work, while Section~\ref{sec:outlook} provides an outlook how the concepts can be applied in operational VGOS sessions. 

\subsection{SNR-based VLBI scheduling} \label{sec:introduction_snr}
The required observation time between two stations can be calculated using
\begin{equation}
    T = \left(\frac{SNR}{\eta \cdot F}\right)^2 \cdot \left(\frac{SEFD_1 \cdot SEFD_2}{rec}\right) \label{eq:t}
\end{equation} with $F$ being the source flux density (source brightness) per band, $SEFD$ being the station system equivalent flux density (station sensitivity) per band, $SNR$ being the target SNR per band, $\eta$ being a constant efficiency factor, and $rec$ representing the recording rate per band.
Thus, given a target $SNR$ and recording rate, the observation time is determined via $SEFD$ and $F$. 
Within a VLBI scan, the required observation time is calculated per baseline (pair of telescopes) and frequency band. 
The minimum over all these observation times per baseline and band determines the final observation time of the given scan. 

While $SEFD$ is measured at most VGOS stations, the main limitation of using an SNR-based scheduling approach for VGOS is that no source flux density ($F$) models are available for VGOS frequencies that are suitable for VLBI scheduling. 
Existing source flux density monitoring campaigns focus on a small subset of sources and use local baselines only, for example, \citet{Varenius2022}, which does not provide suitable results for global VGOS scheduling.

Source flux density is a function of the baseline length and orientation (projected in the direction of the observed source, the so-called UV plane), frequency, and time. 
Currently, almost all IVS schedules are utilizing source flux density models which are part of the sked catalogs\footnote{\url{https://github.com/nvi-inc/sked_catalogs/}}. 
The sked catalog standard supports two types of models: models based on projected baseline length (labeled "B"), where the flux density is defined as a step-function, and elliptical Gaussian models (labeled "M"), where the flux density is defined via the sum of Gaussian components \citep{Vandenberg1997, Schartner2019phd}.
While "B" considers variations solely based on the projected baseline length, "M" allows to consider variations from baseline length and orientation. 
Temporal variations are represented by updating the catalog monthly. 
Frequency-based variations are represented by providing individual models for S- and X-band. 
Since the sked catalog only contains models for S- and X-band, they can not directly be used for VGOS which operates at different frequencies. 

It is to be noted that for operational VLBI scheduling, the calculation of the required observation time contains significant error margins to compensate for imperfections in the models. 
For the interested readers, we provide a more detailed discussion of these error margins in Appendix A. 

% ############
% ### DATA ###
% ############
\section{Data} \label{sec:data}
\subsection{VGOS-R\&D}
The VGOS-R\&D sessions in 2022 were conducted bi-monthly (i.e. one session every second month). 
Each session was individually designed and further discussed and approved by the VGOS Technical Committee (VTC). 
Due to a substantial backlog in VGOS correlation, results from previous sessions were not available before the subsequent sessions were observed. 
Hence, it was not possible to build on previous session results. 
Adjustments made between sessions could only be based on VTC discussions and log files obtained during observations. 

\begin{figure}
    \centering
    \includegraphics[width=.8\textwidth]{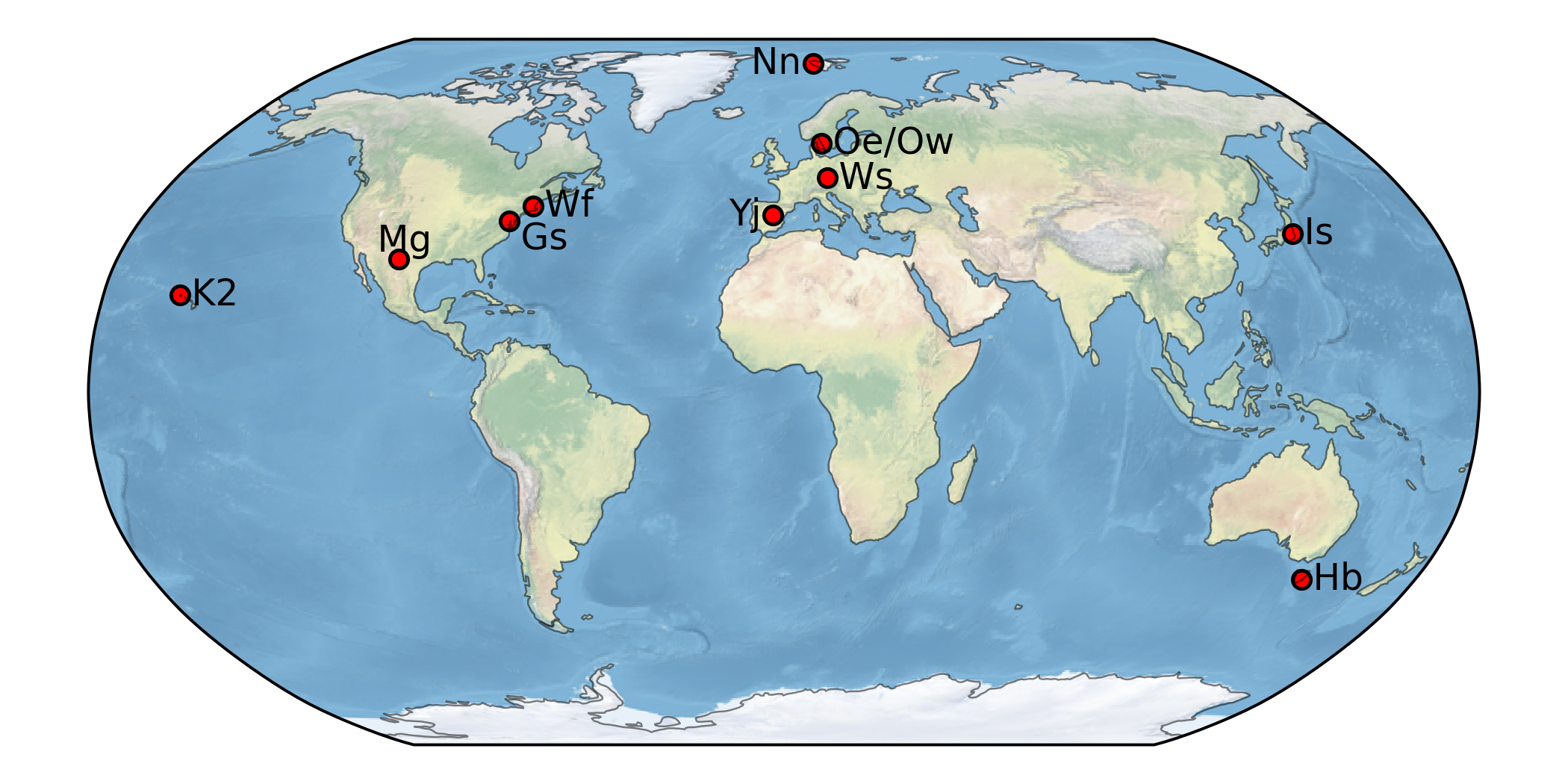}
    \caption{VGOS telescope network: GGAO12M (Gs), HOBART12 (Hb), ISHIOKA (Is), KOKEE12M (K2), MACGO12M (Mg), NYALE13N (Nn), ONSA13NE (Oe), ONSA13SW (Ow), WETTZ13S (Ws), WESTFORD (Wf), RAEGYEB (Yj).}
    \label{fig:worldmap}
\end{figure}

\begin{table}[htb]
    \caption{List of 2022 VGOS-R\&D sessions. 
    A full list of station names can be found in Figure~\ref{fig:worldmap}. 
    Underlined stations were scheduled in tagalong mode. 
    Striked-through stations were included in the schedule but did not observe due to technical issues. 
    }
    \label{tab:vr22_base_stats}
    \centering
    \begin{tabular}{lll}
    \toprule
         Session & Start & Network \\
    \midrule
         VR2201 & 2022-01-20 18:00 & Gs \uline{Hb} K2 Mg Oe Ow Wf Ws \sout{Yj} \\
         VR2202 & 2022-03-17 18:00 & Gs K2 Mg Oe Wf Yj \uline{\sout{Hb}} \uline{\sout{Ow}} \sout{Ws} \\
         VR2203 & 2022-05-19 18:00 & Gs Is K2 Mg Ow Wf Yj \uline{\sout{Oe}} \\
         VR2204 & 2022-07-21 18:00 & Gs \uline{Hb} K2 Mg Oe \uline{Ow} Wf Ws Yj \sout{Is} \\
         VR2205 & 2022-09-15 18:00 & Hb Is K2 Mg \uline{Nn} Oe \uline{Ow} Wf Ws Yj \sout{Gs} \\
         VR2206 & 2022-11-09 18:00 & Hb K2 \uline{Mg} Nn Oe Wf Ws \sout{Gs} \uline{\sout{Ow}} \sout{Yj} \\         
    \bottomrule
    \end{tabular}
\end{table}

Table~\ref{tab:vr22_base_stats} lists the start time and station network of each session, while Figure~\ref{fig:worldmap} depicts the VGOS station network. 
Stations Hb and Nn were initially observed in tagalong mode as they were at this time relatively new and untested, resulting in unstable performance. 
Tagalong mode refers to a scheduling technique, where the observing plan is first generated without the tagalong stations before adding them to the existing schedule. 
This way, losing the tagalong stations will not impact the schedule of the remaining stations. On the other hand, stations observing in tagalone mode cannot contribute their full potential to the network.
Similarly, tagalong mode was occasionally utilized for Oe and Ow due to potential storage limitations affecting the station's ability to observe. 
In the last session, station Mg operated in tagalong mode due to technical issues. 
As presented in Table~\ref{tab:vr22_base_stats}, all sessions, except VR2203, experienced station losses from the core network due to technical problems. 
Consequently, the generated schedule could not be fully executed as intended, leading to limitations in data interpretability to some extent. 

\subsection{VGOS-OPS}
The baseline for comparison is provided by the VGOS-OPS sessions conducted in 2022. 
Specifically, the VGOS-OPS dataset includes 42 sessions, ranging from VO2013 to VO2363 (2022-01-13 to 2022-12-29).
The individual session start times and station network information can be found in the schedule master\footnote{\url{https://ivscc.gsfc.nasa.gov/sessions/2022/}}.

Similar to VGOS-R\&D, the VGOS-OPS sessions suffered from significant station dropout. 
From the original 42 sessions, only 5 could be analyzed with the full station network. 
In the remaining 37 sessions, at least one station did not observe or did not produce useable results. 
Unfortunately, no information regarding station tagalong status is recorded in the available data sources. 

\begin{figure}
    \centering
    \includegraphics[width=\textwidth]{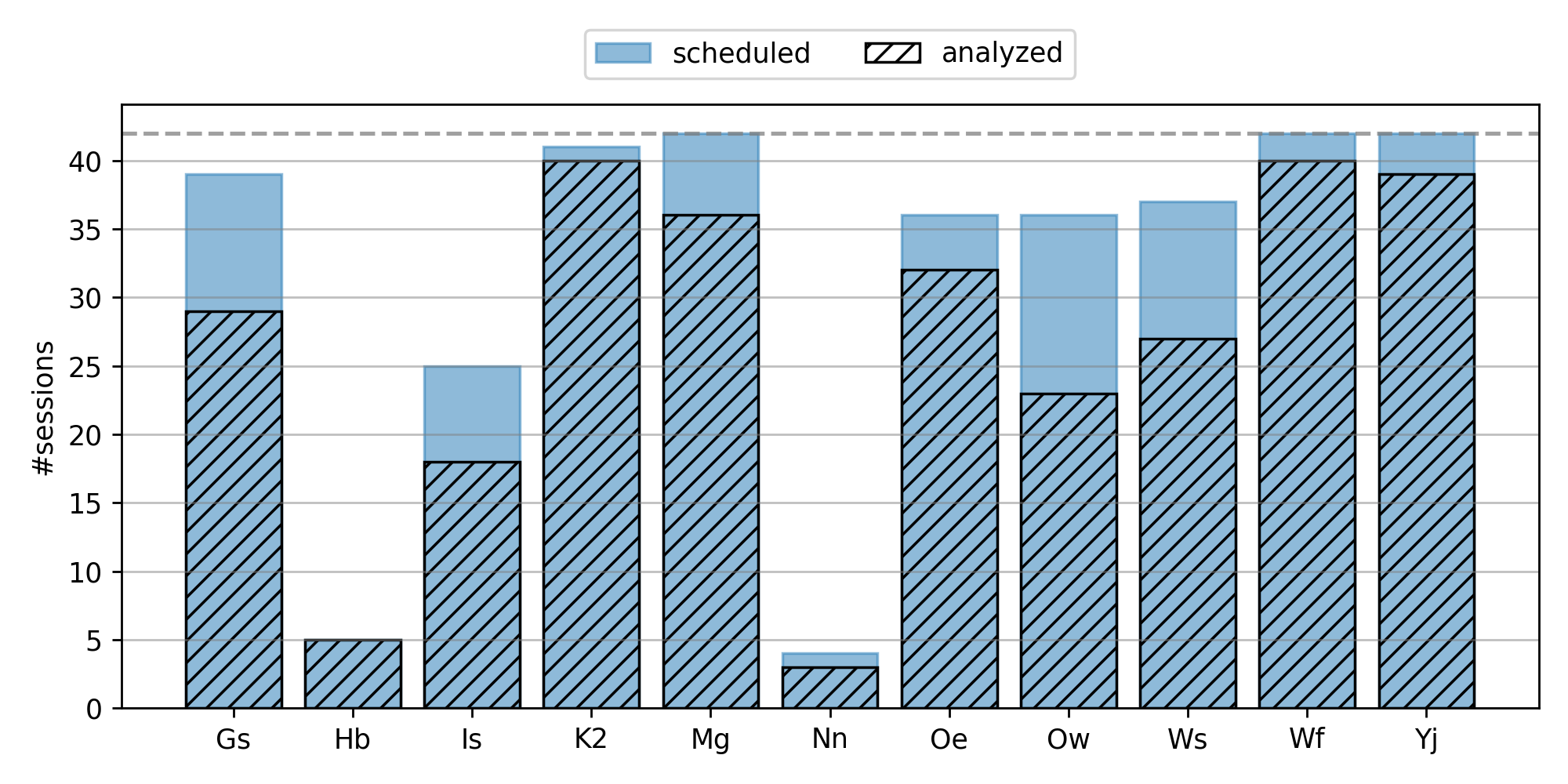}
    \caption{Number of times a station was scheduled and analyzed in a VGOS-OPS session. 
    The dashed gray line marks the total number of VGOS-OPS sessions. }
    \label{fig:sta_vgos_ops}
\end{figure}

Figure~\ref{fig:sta_vgos_ops} depicts how often stations were scheduled and analyzed in the VGOS-OPS sessions. The dashed gray line marks the total number of sessions. 
Within the VGOS-OPS dataset, there is one additional station (KATH12M located in Australia) that participated in one session only. 
Due to this very small sample size and since the station was never observed in the VGOS-R\&D observing program, this station was excluded from the discussion in the following subsections. 

\section{Methodology} \label{sec:method}
The main novelty introduced in the VGOS-R\&D program compared to VGOS-OPS was the introduction of the SNR-based scheduling approach.
While S/X-VLBI sessions already utilized an SNR-driven scheduling strategy for decades, it has not yet been possible for VGOS, in particular, due to a lack of source flux density models at VGOS frequencies. 
In the following, two strategies for deriving the expected source flux density $F$ at VGOS frequencies are discussed that were explored in the VGOS-R\&D program. 
The expected source flux density is then used in calculating the required observing duration using equation (\ref{eq:t}) with a target $SNR$ of around 15, an efficiency factor $\eta$ of around 0.6, and a recording rate $rec$ of \SI{2}{Gbps} per band. 

\subsection{Source flux density estimation} \label{sec:method_snr}
As a first approach, the possibility of a simple inter-/extrapolation from the previously mentioned S/X flux models was tested. 
This was done based on the source spectral index using
\begin{align}
    F(\lambda_i) =& \frac{F_X}{\lambda_X^\alpha} \cdot \lambda^\alpha_i \label{eq:si1}\\
    \alpha =& \frac{\log(F_X/F_S)}{\log(\lambda_X/\lambda_S)}
    \label{eq:si2}
\end{align}
where $\lambda$ stands for the wavelength, $F$ stands for the source flux density and the indices represent the frequencies $S$, $X$, and the target (VGOS) frequency $i$ of bands A--D. 
In the first step, the source flux density is calculated for the projected baseline length and orientation at the S/X frequencies using the sked catalog models. 
Next, equations (\ref{eq:si1}) and (\ref{eq:si2}) are used to calculate the expected source flux density at the VGOS frequencies. 

This approach comes with three major downsides or simplifications. 
First, strictly speaking, an inter-/extrapolation based on the spectral index is only suitable for point-like sources. 
Although the radio sources most commonly observed with VGOS are primarily point-like, this is certainly not the case for all of them. 
Second, in practice, the extrapolations to frequencies outside S/X could become problematic. 
In the case of VGOS, this affects band D. 
Third, errors in the sked S/X catalog are propagated to the VGOS-frequencies. 
It is known that the sked source flux density models have inconsistencies w.r.t. models defined in other catalogs \citep{Lim2022}. 
Here, it is important to understand that the sked models' use case is the calculation of the required observing time. 
This means that in this context, it is necessary to view the source flux density models together with the corresponding $SEFD$ models and for the purpose of solving equation (\ref{eq:t}). 
Potential inconsistencies in the source flux density models can result from assumptions regarding the station $SEFD$ models and vice-versa. 
Similarly, these potential offsets and inconsistencies can also be compensated by the $SEFD$ models for the calculation of the required observing time. 
In practice, the models defined in the sked catalog are used in almost all IVS S/X sessions and represent the state-of-the-art solution.
They are well tested and result in a high observation success rate, potentially also due to the existing high error margins compensating for imperfections. 

The second approach explored in VGOS-R\&D is the use of a newly generated source flux density model at VGOS frequencies, which can be found in the supplementary material attached to the manuscript (in the sked catalog format), as well as in the Appendix B.  
The VGOS source flux density models were generated based on observations from past VGOS-OPS sessions. 
Based on these observations, the catalog includes models for 138 sources at the four VGOS frequencies. 
The models are defined as a step function based on the projected baseline length with a regular stepsize of \SI{1000}{km}. 
This way, they follow the sked catalog conventions and can be supported natively in the scheduling software packages. 
The primary downside of using the new source flux density models is that they only include a limited set of sources and that they were derived using only a small amount of VGOS sessions with limited geometry. 
In particular, the observations lack long north-south baselines, thus, the decision to represent them as circular models based on the projected baseline length instead of elliptical Gaussian models. 
Furthermore, there are only a few observations of sources in the southern hemisphere and none in the deep south. 
Besides, not all stations regularly report their $SEFD$ values in the station logs, further limiting the amount of usable data to derive source flux density models. 
Comparing the VGOS source flux model with the inter-/extrapolation approach, it is evident that there are discrepancies. 
In particular, the new models are more conservative in the reported flux densities, especially for band D. 

\begin{sidewaysfigure}
    \centering
    \includegraphics[width=\textwidth]{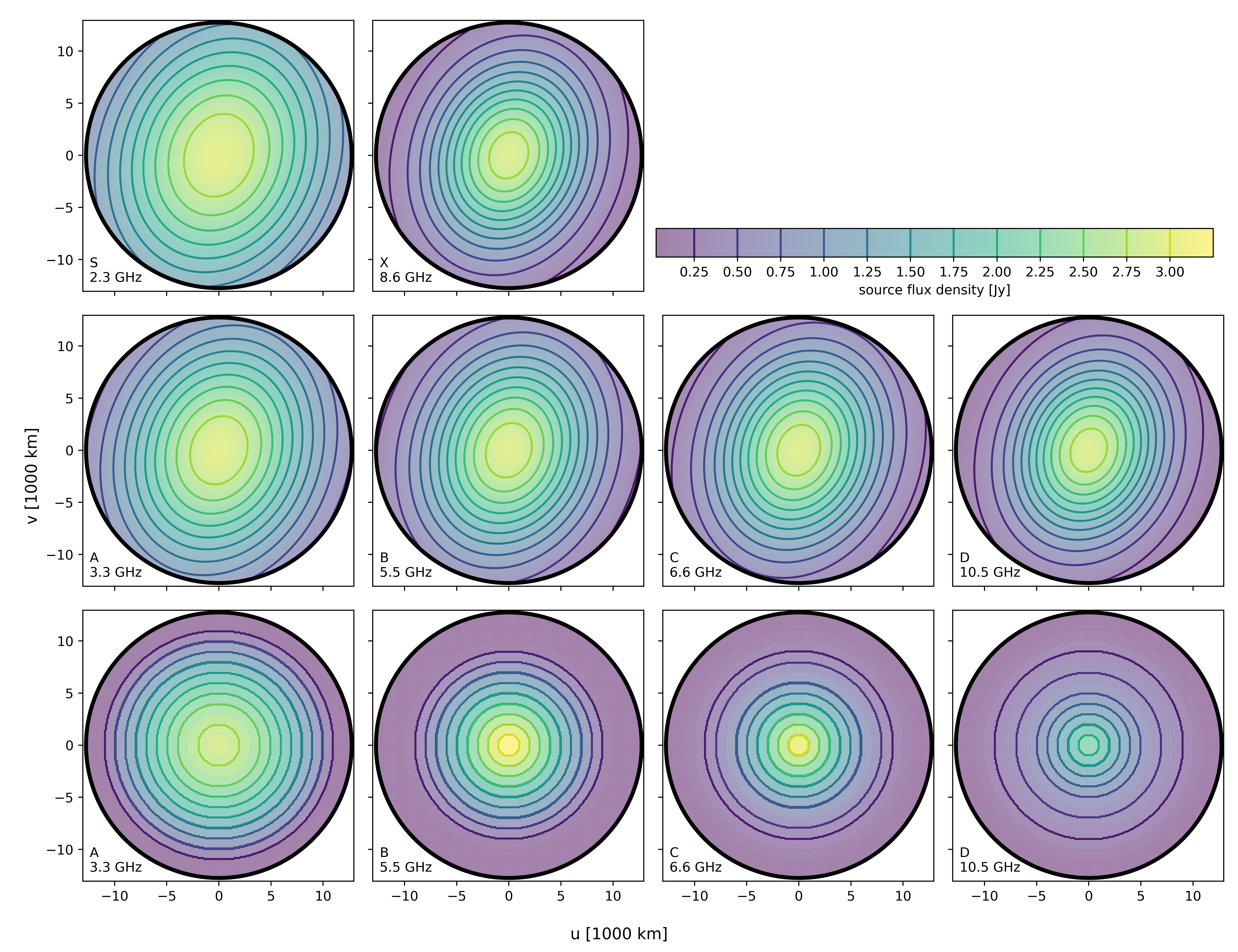}
    \caption{Comparisons of flux density models for source 0552+398. 
    The first row depicts the S/X flux densities reported in the sked catalog. 
    The second row depicts the inter-/extrapolation based on the spectral index. 
    The third row depicts the newly derived VGOS-frequency source flux density models. }
    \label{fig:fluxmodel}
\end{sidewaysfigure}

Figure~\ref{fig:fluxmodel} provides an exemplary depiction of the source flux density model for source 0552+398 to provide a visual example to better understand the concepts. 
The source flux density is color-coded based on the projected baseline length and orientation, defined in the UV plane. 
Note that the original S/X model for source 0552+398, depicted in the first row, is defined based on an elliptical Gaussian model. 
The inter-/extrapolation at VGOS frequencies is depicted in the second row. 
The third row represents the newly generated VGOS frequency source-flux density models, which are defined as a step function based on the projected baseline length. 

\subsection{Scheduling strategy} \label{sec:meth_scheduling}
Unlike VGOS-OPS, the VGOS-R\&D schedules were generated using VieSched++ \citep{Schartner2019}. 
The general scheduling concept followed the approaches described in \citet{Schartner2020a} and \citet{Schartner2021b}.

One requirement posed by the SNR-driven observation strategy is the requirement of dedicated calibration scans. 
In the VGOS-OPS sessions, where each scan is observed for 30 seconds straight, enough scans reach high enough SNR to naturally provide good calibration capabilities. 
However, due to the reduced observation time in VGOS-R\&D sessions, resulting in lower SNR, this is not necessarily the case anymore and dedicated calibration scans should be included in the session. 
Based on feedback within the VTC, as well as from the correlator staff [M. Titus, personal communication, 2022], these calibration scans were included every 1-2 hours.
Within the 2022 VGOS-R\&D sessions, the observation time of calibration scans was set to 60 seconds. 
The calibration scans were selected solely based on source visibility and acquired SNR. 

Another reason to include 60-second long calibration scans in VGOS-R\&D sessions, unrelated to the SNR-based observations strategy, was the inclusion of some new and not yet fully validated stations (Hb and Nn), where these high SNR scans provided valuable insight into the station performance. 

The objective of the VGOS-R\&D sessions was to evaluate the suitability of both previously discussed SNR-based scheduling approaches despite their assumptions and limitations. 
The inter-/extrapolation approach based on the sked catalog was utilized in sessions VR2201 and VR2202 while the new VGOS source flux density models were used in VR2203--VR2205. 
In VR2206, the majority of sources used the VGOS source flux density models while a few additional sources were added in the session that were not included in the new catalog. 
For these sources, the inter-/extrapolation approach based on the sked catalog was used. 
For the SNR-based scheduling strategy, a minimum observing time of 7 seconds and a maximum observing time of 20 seconds were used, except for VR2204, where the minimum and maximum were reduced by 2 seconds each. 
The resulting average observation time of scans within VGOS-R\&D is 10 seconds. 

\begin{figure}
    \centering
    \includegraphics[width=.7\textwidth]{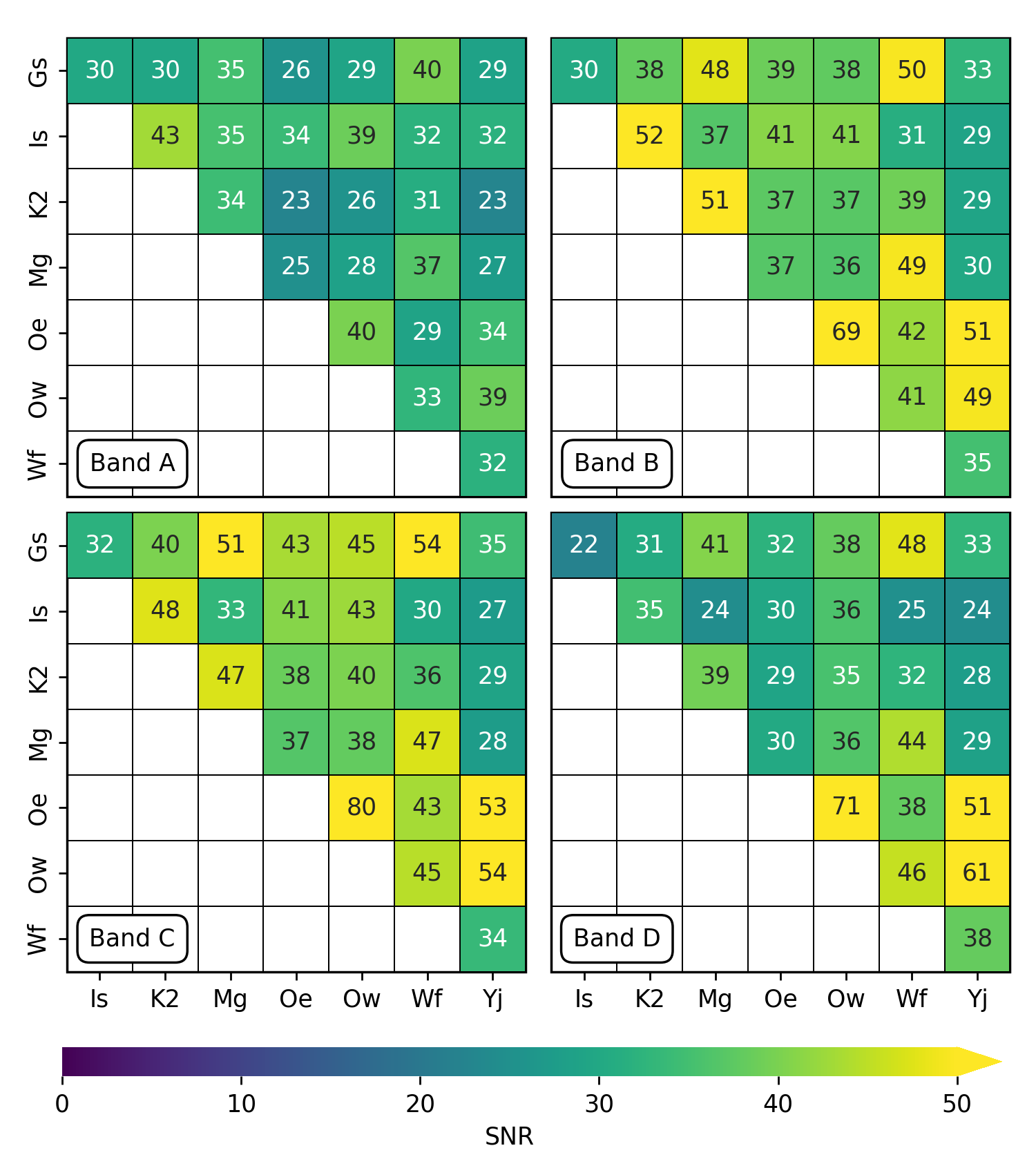}
    \caption{Theoretical average SNR per baseline and band for session VR2203. Results for all sessions can be found in Appendix C. }
    \label{fig:theoretical_snr}
\end{figure}

As discussed in Section~\ref{sec:introduction_snr}, the required observation duration is calculated per band and baseline. 
Consequently, only one band of one observation has a theoretical SNR close to the target SNR, while all other remaining bands and observations have a higher theoretical SNR. 
As an example, Figure~\ref{fig:theoretical_snr} depicts the average SNR per band and baseline for one exemplary session VR2203 (other sessions can be found in Appendix C). 
In this case, the target SNR was set to 15 for all bands and on all baselines. 
However, the average SNR is significantly larger than 15 due to the previously discussed facts, especially for bands B and C where telescopes typically have higher sensitivity compared to bands A and D. 

\subsection{Overhead times} \label{sec:overhead}
Besides utilizing an SNR-based scheduling strategy, which drives the majority of the increase in the number of obtained scans per hour, a careful evaluation of existing overhead times in VGOS operations was executed for the VGOS-R\&D program which is explained in the following. 

For each scan, the telescope has to execute a variety of steps. 
Simplified speaking, it has to slew to the radio source that will be observed (slewing), wait for all other telescopes to finish slewing (idle time), and observe the radio source (observing time).
While slewing and idle times are variable and depend on the telescope properties, the observing time is fixed to 30 seconds within VGOS-OPS sessions as discussed earlier. 
Additionally, before each observation, some time is reserved to set up the upcoming observation and perform some calibrations. 
Within the VGOS-OPS session, this calibration time is fixed to four seconds per scan. 
Finally, there are two reasons to account for some additional overhead time during the scheduling process. 
First, for each scan, a fixed constant amount of overhead time is reserved, intending to reflect the execution time of field system commands. 
Within VGOS-OPS sessions, this constant overhead time is set to four seconds. 
Second, due to limitations in the recording hardware used in VGOS-OPS session, an additional overhead time in the same length of the observing time is required. 
It serves to guarantee that all recorded data is properly written to hard drives before the next scan starts. 
Here, it is to note that during this overhead time, the telescopes can already slew to the next scan. 
Thus, while generating the schedules, it can be seen as a constraint on the slewing time instead of an additional overhead time. 

Based on an evaluation of these times for VGOS-R\&D, it was found that the four-second-long overhead time for executing field system commands is not required and can be removed. 
Additionally, applying some slight modifications to the telescope procedures executed during the calibration phase could reduce it to only two seconds. 
Finally, using a second recording module eliminated the additional overhead time to ensure that all recorded data is properly written to hard drives before the next scan starts. 
However, it has to be noted that for the VGOS-R\&D sessions, this step is in general far less significant compared to VGOS-OPS sessions. 
The reason is that the observing time was greatly reduced to around 10 seconds as discussed in \ref{sec:meth_scheduling} which consequently also reduced the required overhead time which often became shorter than the simultaneously applied slewing time. 
For example, when generating the same schedule for VR2201 with the same scheduling strategy except for one version assuming recording on one module only (and thus requiring the overhead time) while the other version assumes recording on two modules (and thus does not require the overhead time) the resulting increase in terms of number of scans in the second version is only \SI{0.4}{\%} compared to the first version, while the number of observations is increased by \SI{3}{\%} only. 

\begin{figure}
    \centering
    \includegraphics[width=\textwidth]{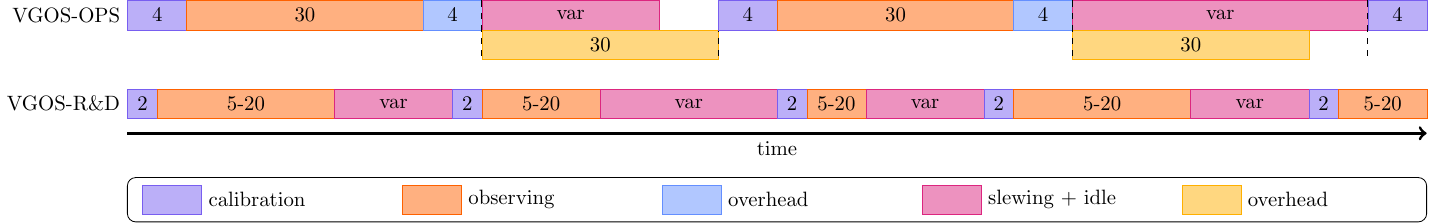}
    \caption{Sketch of observing sequence including overhead times for VGOS-OPS and VGOS-R\&D (not at scale). The numbers represents a duration in seconds. "var" means variable time. }
    \label{fig:sketch}
\end{figure}

Figure~\ref{fig:sketch} provides a visual sketch comparing the observation approaches of VGOS-OPS and VGOS-R\&D. 
For simplicity, the first overhead time, reserved to execute field system commands, is displayed before the slewing starts. 
The second overhead time, reserved to ensure that all recorded data is safely stored on disk, is executed simultaneously with the slewing and idle time. 

\subsection{Simulation strategy} \label{sec:methodology_simulations}
To determine the expected precision of the SNR-based scheduling strategy, Monte-Carlo simulations were executed to derive the expected precision of the estimated parameters based on their repeatability error. 
The simulation strategy follows state-of-the-art procedures \citep{Pany2011}.
It considered three major error sources: tropospheric turbulences, clock drifts, and random measurement errors. 
The tropospheric turbulence was simulated using spatial and temporal correlations with an average refractive index structure constant $C_n$ of $\SI{1.8e-7}{m^{-1/3}}$ and an effective wet tropospheric height of $\SI{2000}{m}$, and a wind velocity of \SI{8}{m/s} \citep{Nilsson2007, Nilsson2010}. 
Clock drifts were simulated as an integrated random walk with an Allan standard deviation of $\SI{1e-14}{s}$ over 50 minutes. 
Finally, normally distributed measurement errors with a standard deviation of 4 picoseconds were added. 
The simulations were analyzed with a standard least squares adjustment, estimating station coordinates, clock parameters (a quadratic polynomial, as well as piece-wise linear offsets (PWLO) every 30 minutes, constrained with \SI{1.3}{cm}), tropospheric parameters (PWLO zenith wet delay every 15 minutes with constraints of \SI{1.5}{cm}, as well as PWLO north-south and east-west gradients with constraints of \SI{0.05}{cm}, and all five Earth orientation parameters (EOP) as tightly constrained (\SI{0.1}{\micro as}) PWLO at session start and session end, effectively delivering one offset estimate. 
All stations were considered in a no-net-rotation, no-net-translation datum definition. 

Each session was simulated and analyzed 1000 times with varying realizations of the simulated error sources to allow for a robust calculation of the Monte-Carlo repeatability error (rep) of the estimated geodetic parameters, calculated via their standard deviation. 
Thus, the repeatability error represents the expected precision.

\subsection{Analysis strategy} \label{sec:methodology_analysis}
While the primary focus of this work is to provide a proof of concept of greatly increasing the number of scans per session, an outlook w.r.t. the geodetic performance is also provided within the manuscript. 

The geodetic performance is validated based on two data sources:
First, based on the official IVS analysis results, obtained via the $\nu$-Solve package \citep{Bolotin2014}. 
Second, based on our analysis, derived from processing group delays provided as databases via IVS Data Centers using the Vienna VLBI and Satellite Software \citep[VieVS; ][]{Boehm2018}.

In our analysis, the a priori group delays were modeled as described in \citet{Krasna2023a}. 
Station position time series were obtained from solutions where the source coordinates were fixed to the ICRF3 \citep{Charlot2020}, while the station coordinates, as well as datum definition, were based on a priori information from the ITRF2020 \citep{Altamimi2023}. 
Tropospheric parameters, including ZWD and gradients, were estimated every 30 minutes with relative constraints of \SI{1.5}{cm} and \SI{0.5}{mm} between the piece-wise linear offsets PWLO, respectively. 

In terms of EOP, polar motion and UT1-UTC parameters were estimated as PWLO every 24 hours at 0~UTC with relative constraints of \SI{10}{mas}.
Since the session starts at 17 UTC, we report the estimates at 0 UTC during the session and at 0 UTC on the day after the session.   
Celestial pole offsets were estimated as 24-hour PWLO with tight constraints of \SI{0.1}{\micro as}, effectively delivering one offset at the mid-epoch of the session, approximately around 5 UTC.
The estimated EOP are consistent with the state-of-the-art global solution VIE2022 \citep{Krasna2023} which produced an updated TRF and CRF including the most recent sessions. 
% The adjustment involved 7308 24h IVS S/X and VGOS sessions from 1979.6 until 2023.0. 

% ###############
% ### RESULTS ###
% ###############
\section{Results and Discussion} \label{sec:results}
\subsection{Scheduling statistics}

\begin{figure}[htb]
    \centering
    \includegraphics[width=\textwidth]{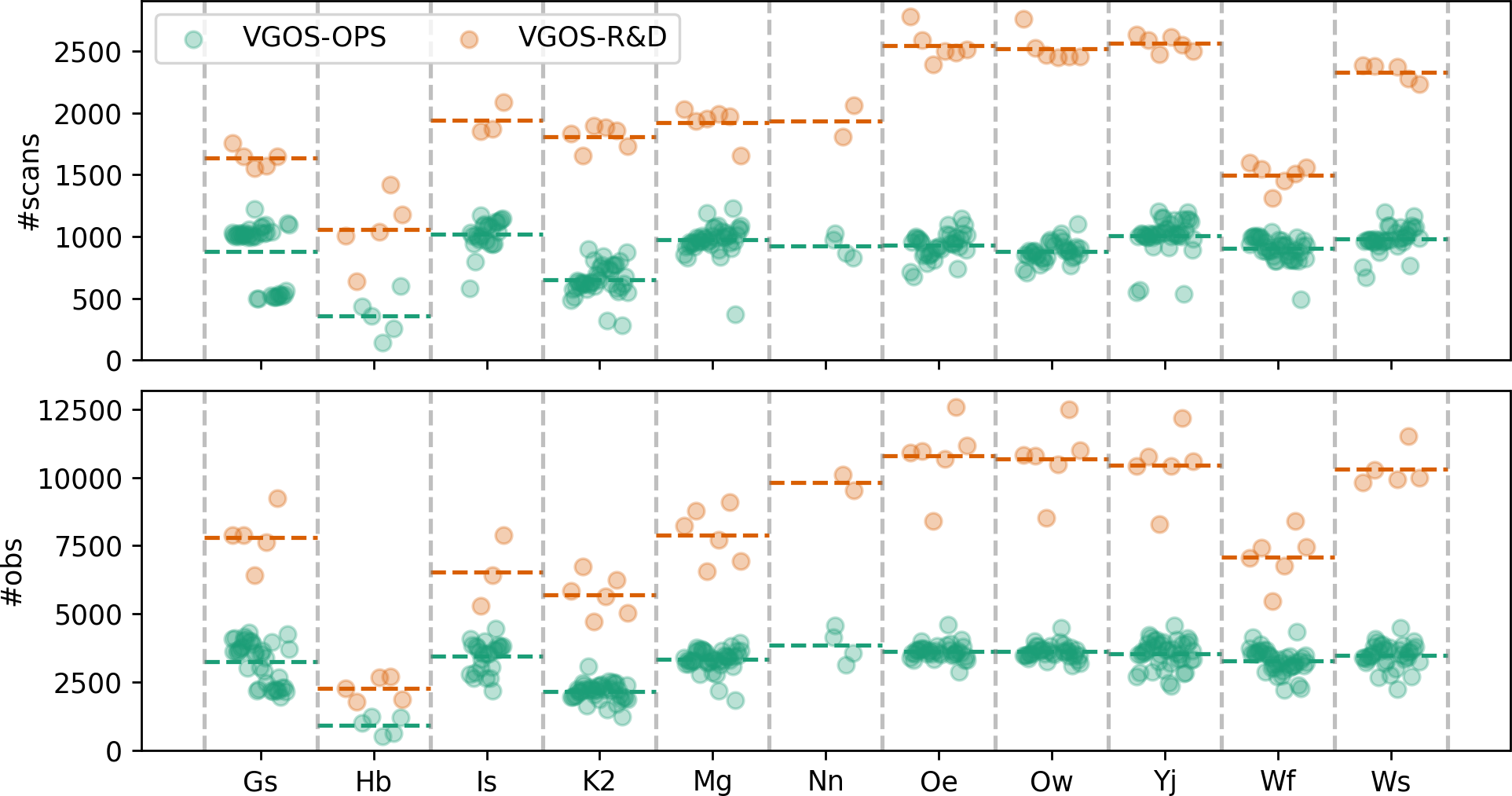}
    \caption{Scheduling statistics; Top: number of scans per station; Bottom: number of observations per station.
    The markers represent individual sessions while the dashed lines depict the average. 
    VGOS-R\&D sessions are depicted in orange while VGOS-OPS sessions are depicted in turquoise. }
    \label{fig:stats}
\end{figure}

Figure~\ref{fig:stats} presents the number of scans and the number of observations of each station per session. 
The dashed line indicates the average number.
Considering VGOS-R\&D relative to VGOS-OPS, the number of scans per station, as well as the number of observations per station, is increased substantially. 
The increases are by a factor of 2.3 and 2.6 respectively. 

\begin{figure}[htb]
    \centering
    \includegraphics[width=\textwidth]{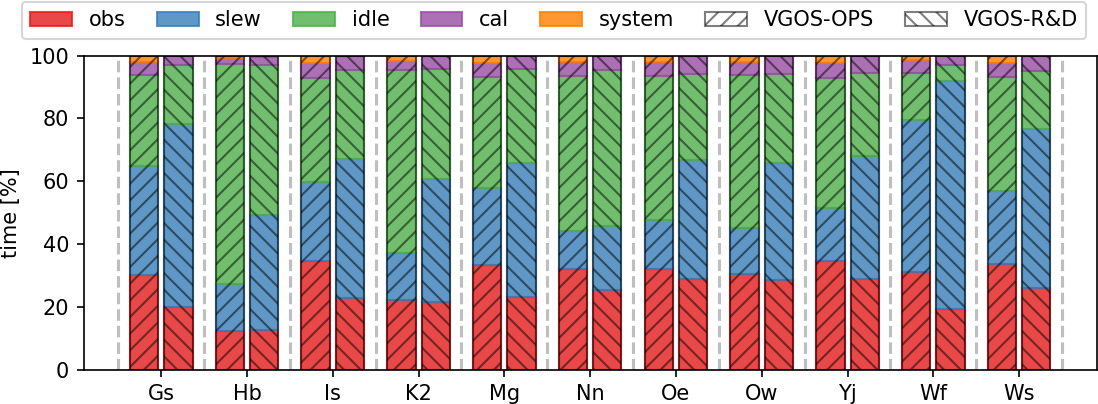}
    \caption{Station by station time distribution of activities; obs: observation time, slew: slewing time, idle: station idling, cal: calibration time, system: time for field-system commands. 
    The VGOS-OPS sessions are depicted in the left columns (with /-hash marks) while the VGOS-R\&D sessions are depicted in the right columns (with \textbackslash -hash marks). }
    \label{fig:time}
\end{figure}

Figure~\ref{fig:time} depicts, on a station-by-station basis, the distribution of time spent in different activities, e.g. observing (obs), slewing (slew), idling (idle), calibration (cal), and execution of field-system commands (system). 
In this context, it is to note that the previously mentioned overhead time intended to ensure that the recorded data is stored safely on disk is not represented since it occurs simultaneously with slewing and idling (see Section~\ref{sec:overhead}). 

The first noteworthy conclusion is that the VGOS-R\&D sessions have \SI{20}{\%} less observing time compared to VGOS-OPS, although 2.3 more scans were observed as discussed previously. 
This is due to the SNR-based scheduling algorithm with an average observing time of 10 seconds per scan compared to the 30 seconds per scan in VGOS-OPS. 
The reduction in observing time also implies less data transfer and fewer bits to be correlated, which is important because data transfer and correlation are the current operational bottlenecks of today's VGOS observations. 
Next, one can see that the slewing time is increased by a factor of 2. 
This indicates that the atmospheric sampling is improved since longer slewing times mean that more different azimuth and elevation angles are observed.
Finally, one can see that there is still significant idle time left. 
However, idle time is always correlated with slewing time since the stations have to wait for the slowest station to finish slewing before starting observations. 
In the VGOS network, the slowest station is Wf which has almost no idle time.
Wf has a slewing rate of 200 degrees per minute in azimuth and 120 degrees per minute in elevation compared to the 720 degrees per minute in azimuth and 360 degrees per minute in elevation of most other VGOS stations. 

\newpage
\subsection{Successful observations} \label{sec:results_success}
\begin{figure}[H]
    \centering
    \includegraphics[width=1\textwidth]{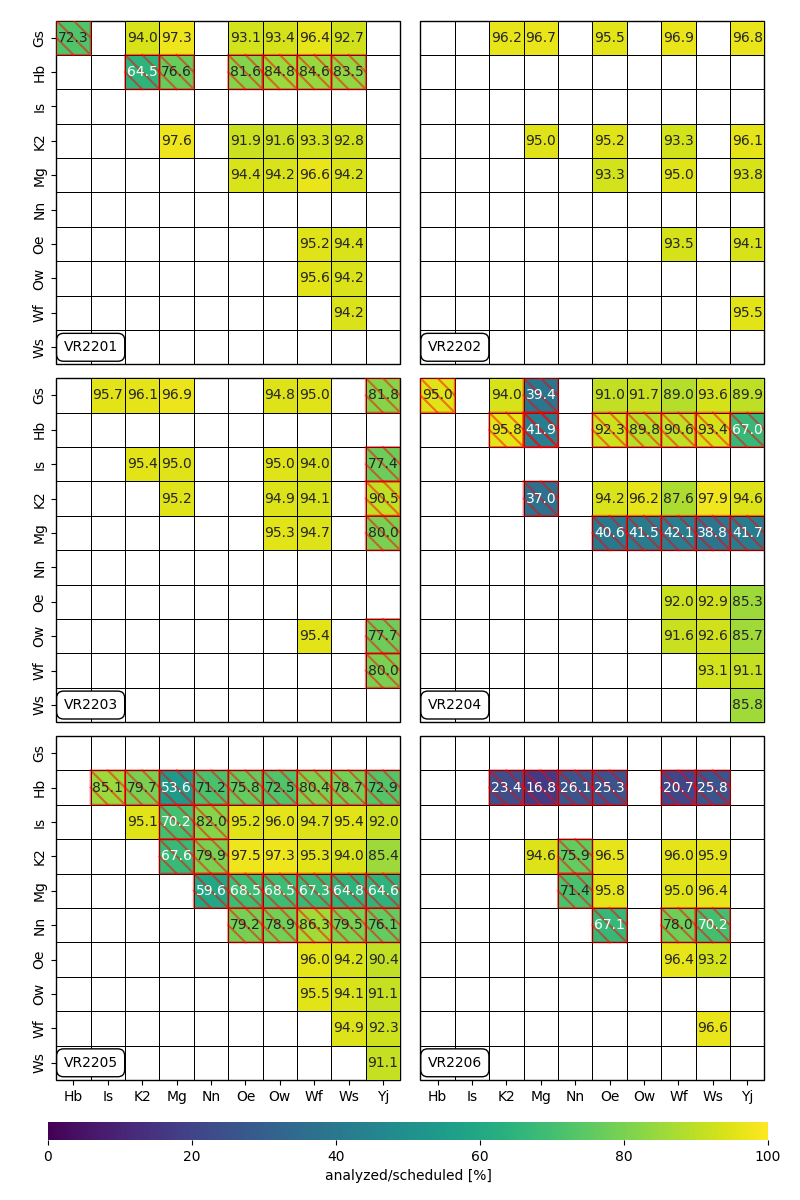}
    \caption{Percentage of successful observations per baseline. 
    The percentage is color-coded and explicitly entered in each cell. 
    White cells represent baselines that have not been observed. 
    Cells with red hatching are not representative since they either belong to the new and not yet fully validated stations Hb and Nn, or since these stations suffered from observation loss due to unrelated technical problems. }
    \label{fig:obs_success}
\end{figure}

The main research question of the VGOS-R\&D sessions was the feasibility of the short, SNR-based observation times. 
We evaluate the SNR-based scheduling approaches by examining the percentage of successful observations. 
In this context, we define a successful observation as an observation used in geodetic analysis. 
Figure~\ref{fig:obs_success} overviews the percentage of successful observations per baseline.
The values were extracted from the official IVS analysis reports. 
Based on this analysis, it can be seen that many baselines have a success rate of more than \SI{90}{\%} which is similar to the success rate of VGOS-OPS sessions. 
Outliers are observations including the, at the time of the sessions, new and not fully validated stations Hb and Nn. 
Furthermore, in VR2203, observations with station Yj had a lower success rate which can be explained by unrelated technical problems experienced at the station. 
Finally, there is a lower success rate at observations including Mg in VR2204 and VR2205. 
In both cases, it can be explained by the station missing approximately \SI{50}{\%} of the session in VR2204 and approximately \SI{20}{\%} of the session in VR2205 due to unrelated problems at the station. 
More details regarding the telescope issues can be found in the analysis reports uploaded at the IVS Data Centers\footnote{\url{https://cddis.nasa.gov/archive/vlbi/ivsdata/aux/2022}}. 
Thus, all of these reduced success rates can be explained by technical problems and are not related to the SNR-based observation strategy. 

\begin{table}
    \raggedleft
    \captionsetup{width=\textwidth}
    \caption{Average observation success rate excluding stations with unrelated technical problems. The first column lists the session name. The second column lists the utilized source flux density modeling approach for the SNR-based scheduling. 
    "interp" stands for an inter-/extrapolation based on the sked S/X catalog, while "models" stands for the newly derived source flux density models at VGOS frequencies. 
    VR2206 used primarily the "models" approach but also some sources with the "interp" approach. The last column lists the percentage of successful observations.}
    \label{tab:successrate}
    % faking a column at the left/right to get the caption to span the full text width :-) 
    \begin{tabularx}{\textwidth}{X l l r X} 
         \cmidrule{2-4}
         & Session & SNR-approach & [\%] &\\
         \cmidrule{2-4}
         & VR2201 & interp & 94.6 &\\
         & VR2202 & interp & 95.1 &\\
         & VR2203 & models & 95.2 &\\
         & VR2204 & models & 90.1 &\\
         & VR2205 & models & 93.4 &\\
         & VR2205 & mixed & 95.3 &\\
         \cmidrule{2-4}
    \end{tabularx}
\end{table}
Table~\ref{tab:successrate} lists the average success rate per session excluding the stations with unrelated technical problems. 
Both approaches, the inter-/extrapolation of the S/X flux information as well as the use of the newly generated source VGOS frequency flux density catalog provided high success rates of over \SI{90}{\%}. 
This proves that the SNR-based observation times are technically possible and that the percentage of successful observations using those approaches is feasible. 
When keeping in mind that VGOS-R\&D scheduled a factor of 2.6 times more observations, one can conclude that the total yield of usable observations for analysis is significantly improved. 

Since there is a difference between the newly derived source flux density models and the inter-/extrapolation approach based on the sked S/X catalog, but no significant difference in the success rates of both approaches, these results hint that most likely the error margins discussed in Appendix A compensate for any imperfections in the models. 
Consequently, by improving station SEFD monitoring and modelling, as well as source flux density monitoring and modelling, the error margins could theoretically be reduced and even shorter observation times might be feasible. 

\subsection{Signal-to-noise ratios}
While the primary metric to evaluate the SNR-based scheduling approach was the percentage of successful observations, studying the observed SNRs per band can also be compared with the predicted SNRs per band to evaluate the accuracy of the VGOS models. 

Unfortunately, information regarding the observed SNRs was not available at the time VGOS-R\&D sessions were scheduled due to significant delays in the correlation of the sessions and in developing the methodologies required to perform SNR analysis. 
Thus, they could not be used to tune subsequent sessions and were only computed after all sessions had already been observed.

Following \citet{Corey22}, the single-band SNRs are reconstructed from the total SNR using equation (\ref{eq:snr_band}) since the observed SNRs per band are not stored and thus available for analysis. 
The reconstruction error is assumed to be very small (typically within \SI{1}{\%}) and is negligible compared to the uncertainties in the theoretical SNR calculation and underlying models. 
\begin{equation}
    SNR_{band} = \frac{SNR_{tot}}{amp_{tot}} \cdot \frac{\sum_{n=1}^N V_n}{\sqrt{M \cdot N}}
    \label{eq:snr_band}
\end{equation}
All variables required in (\ref{eq:snr_band}) are stored in the session vgosDB files, publicly available via the IVS data centers.
$SNR_{tot}$ is the combined four-band SNR, $amp_{tot}$ is the coherent average fringe amplitude for the combined four bands, $V_i$ is the amplitude of the complex fringe visibility of a single channel $i$, $N$ represents the number of channels within a given band, and $M$ represents the total number of channels. 

\begin{figure}
    \centering
    \includegraphics[width=.7\textwidth]{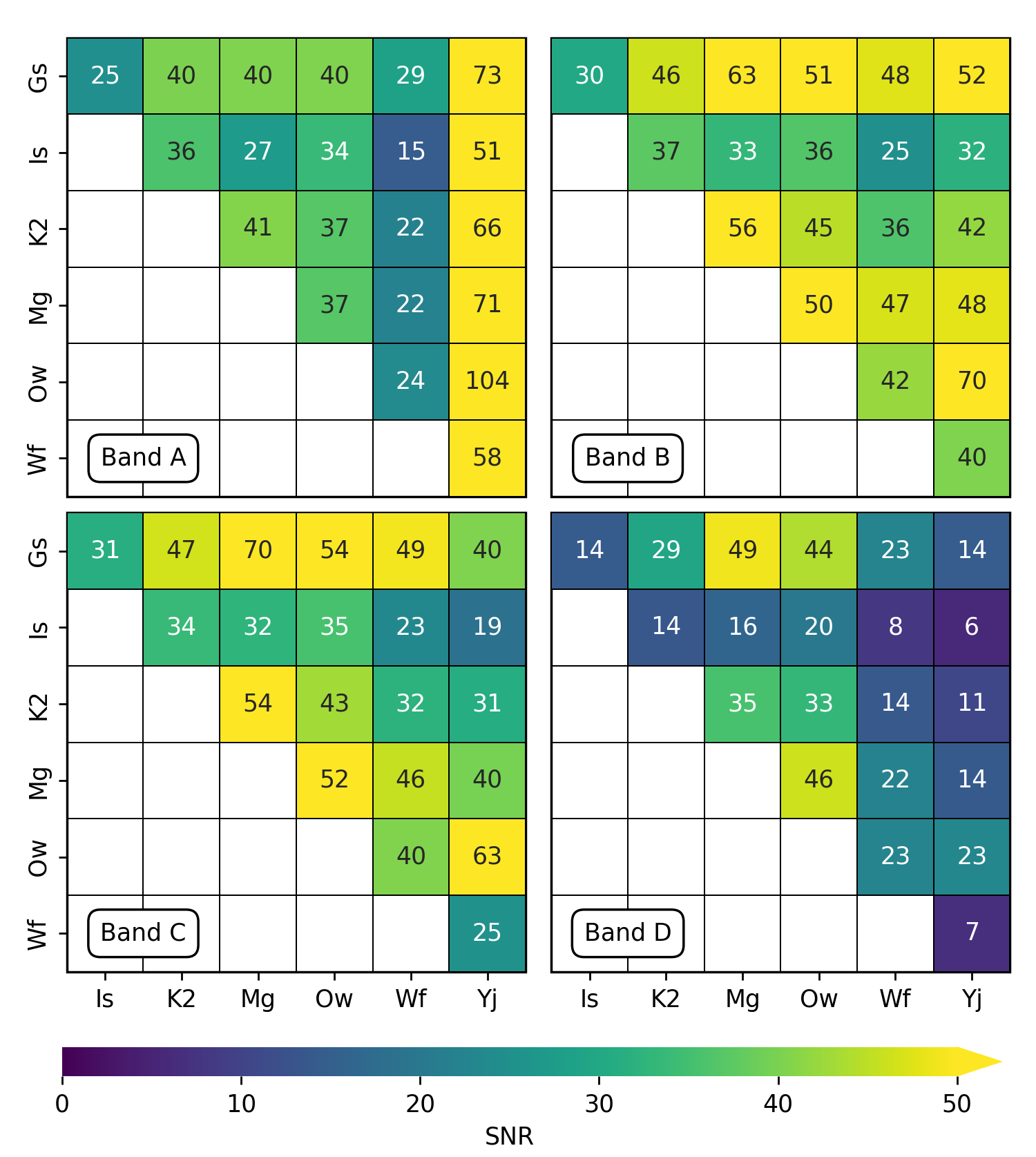}
    \caption{Average reconstructed SNR per baseline and band for session VR2203. 
    Results for all sessions can be found in Appendix D.}
    \label{fig:snr_obs_vr2203}
\end{figure}

Figure~\ref{fig:snr_obs_vr2203} lists the average reconstructed single-band SNRs per baseline for one exemplary session (VR2203), similar to Figure~\ref{fig:theoretical_snr}, where the theoretical SNRs are depicted. 
Results from other sessions can be found in Appendix D. 
It can be seen that in this session, the lowest SNRs are observed in band D. 
Furthermore, some station-dependent effects are visible, like a reduced SNR for observations including Wf in band A, which agrees with analyst comments on the same issue. 

\begin{figure}
    \centering
    \includegraphics[width=.9\textwidth]{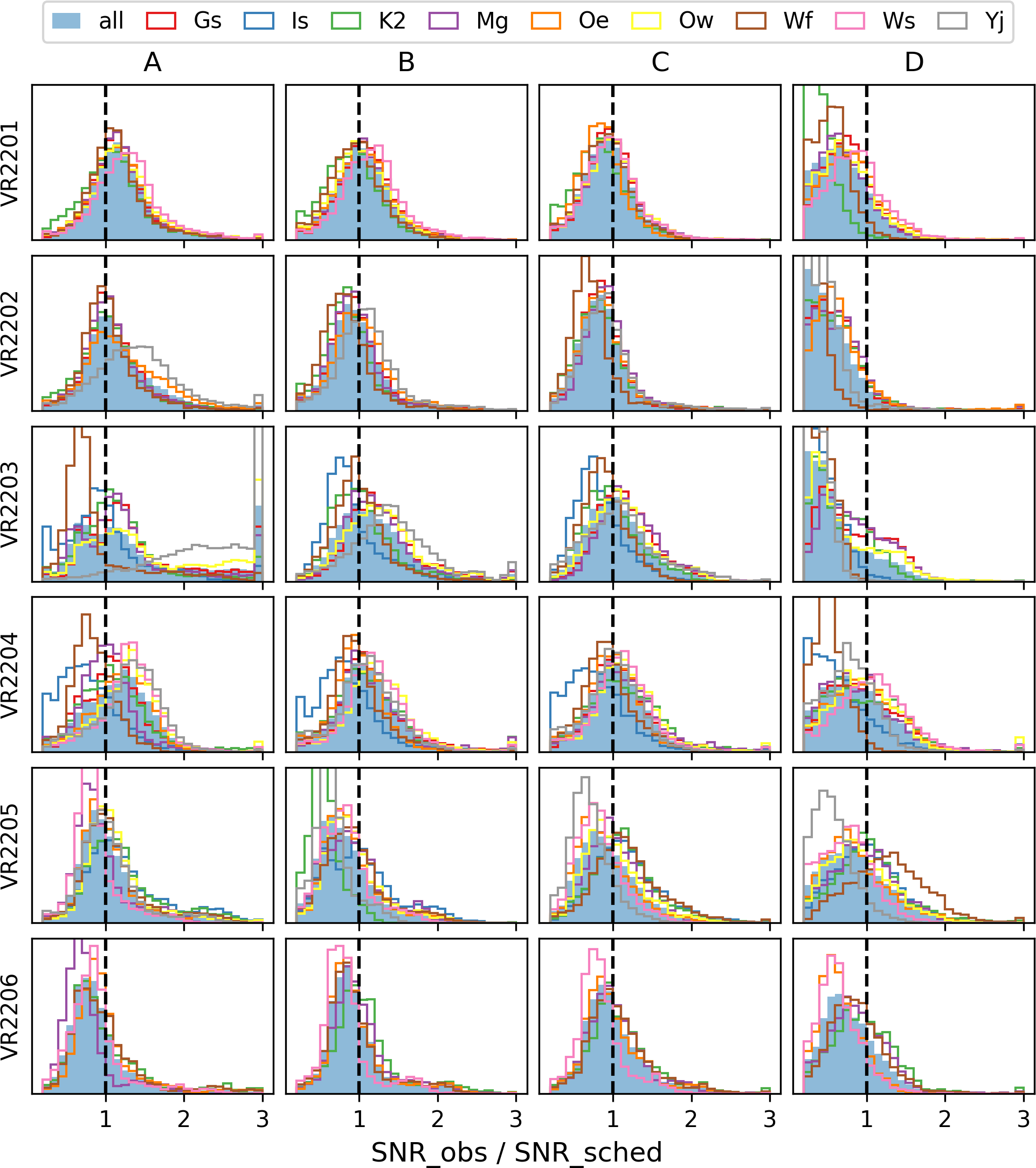}
    \caption{Ratios between reconstructed SNR (SNR\_obs) and predicted SNR (SNR\_sched) per session and band. The average distribution is depicted as a solid blue histogram. Distributions per station are depicted as colored lines. Rows depict individual sessions while columns depict individual bands. The ratio of one is highlighted as a dashed black line. }
    \label{fig:snr_ratios}
\end{figure}

Figure~\ref{fig:snr_ratios} depicts the distribution of the ratios between the reconstructed SNR (SNR\_obs) and the predicted SNR (SNR\_sched). 
A dashed black line highlights the ratio of 1.00 which represents perfect agreement. 
Ratios $<1.00$ mark observations where the models used in the calculation of the theoretical SNR were too optimistic, while ratios $>1.00$ depict cases where the models were too conservative. 
Observations with stations Hb and Nn are excluded since these stations were at the time not yet fully validated and no information regarding their sensitivity was available. 
Based on Figure~\ref{fig:snr_ratios}, several conclusions can be drawn. 

First, in many cases, the predicted SNRs are too optimistic. 
The overly optimistic SNRs are most pronounced in band D of the early VGOS-R\&D sessions (VR2201--VR2203). 
In general, stations tend to have a lower sensitivity in band D. 
Since the actual SEFD was not monitored at all stations at the time the VGOS-R\&D sessions were observed, assumptions have to be made, which might have been too optimistic. 
Furthermore, the extrapolation of the source flux densities in VR2201 and VR2202 (see Section~\ref{sec:meth_scheduling}) might have led to higher expected flux densities than observed in reality. 
As further discussed in Section~\ref{sec:method_snr}, comparisons between the inter-/extrapolation-based source flux density models and the newly derived ones had the highest disagreement in band D as well, with the new models reporting lower flux densities. 
This might also explain why the ratios in the later session (VR2203 -- VR2206) are close to 1.00. 

%Second, it is evident that in particular, the ratios are worst for observations including station Hb. 
%This is not surprising since station Hb was at that time participating in VGOS sessions for the first time. 
%Consequently, it was not yet fully operational and validated. 
%Furthermore, it tended to have the lowest sensitivity. 
%Combined with them being scheduled in tagalong mode, this resulted in the lowest predicted SNR to begin with. 

Second, although the observed SNRs are in many cases quite low, the VGOS processing pipelines still manage to extract useable observations in many cases. 
Thus, they seem to be quite robust. 

Third, the spread of the SNR ratios indicates that there is still significant room for improving the VGOS models to predict more accurate SNRs. 
More resources and research are needed to tackle the remaining shortcomings. 

Finally, although the SNR modeling shows room for improvement and is sometimes too optimistic, the error margins (e.g. by targeting a higher SNR as required) compensate for the inaccuracies and lead to a high observation success rate as discussed in Section~\ref{sec:results_success}. 
Furthermore, it is important to discuss these results in context with the original objective of the VGOS-R\&D program. 
The original research objective was to significantly increase the number of scans compared to the current state-of-the-art approaches.
Thus, it can be concluded that the combination of the modeling approaches, error margins, and observation duration between \SIrange{5}{20}{s} used in VGOS-R\&D are sufficient to extract usable observations. 
Consequently, observing every scan for 30 seconds straight as done in VGOS-OPS is not strictly necessary.  

\subsection{Simulation results}
\begin{figure}
    \centering
    \includegraphics[width=\textwidth]{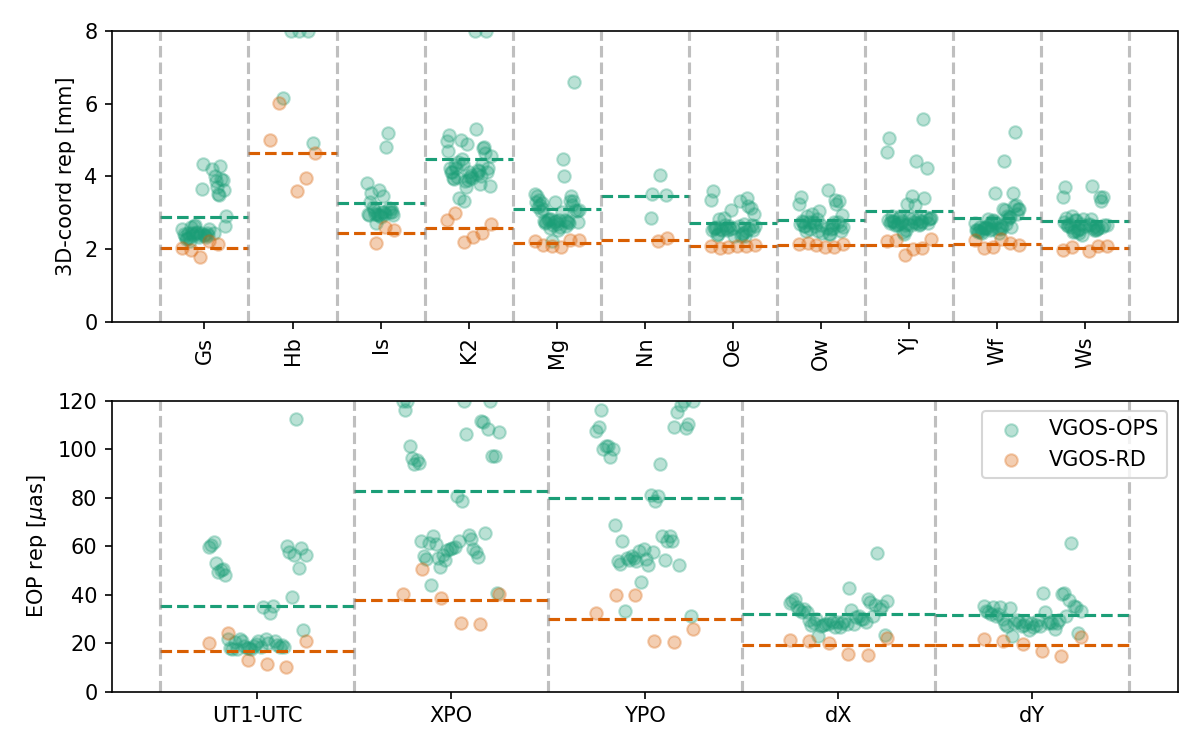}
    \caption{Top: simulated repeatability error of 3D station coordinates; Bottom: simulated repeatability error of EOP. 
    The markers represent individual sessions while the dashed lines depict the average. 
    VGOS-R\&D results are depicted in orange while VGOS-OPS results are depicted in turquoise. 
    3D-coordinate repeatability errors larger than \SI{8}{mm} are displayed at \SI{8}{mm} while EOP repeatability errors larger than \SI{120}{\mu as} are displayed at \SI{120}{\mu as}}.
    \label{fig:simulations}
\end{figure}

To determine the expected precision of the geodetic parameters, simulations of all VGOS-R\&D and VGOS-OPS sessions were conducted as discussed in Section~\ref{sec:methodology_simulations}.
Figure~\ref{fig:simulations} depicts the simulated geodetic precision expressed via the repeatability errors (rep) from the Monte-Carlo simulation. 

Comparing the 3D station coordinate repeatability errors $\sqrt{rep_X^2 + rep_Y^2 + rep_Z^2}$ of VGOS-OPS and VGOS-R\&D, it can be seen that the expected precision is higher for VGOS-R\&D sessions, likely explained via the increased number of scans per station. 
Except for station Hb, which is poorly integrated due to its remote location, especially in VGOS-OPS sessions, the average reduction in the repeatabilities of VGOS-R\&D compared to VGOS-OPS is \SI{50}{\%}. 
For the five EOP, the average improvement in repeatability error is also \SI{50}{\%}, \SI{50}{\%} for UT1-UTC, \SI{60}{\%} for polar motion (XPO, YPO), and \SI{40}{\%} for the nutation parameters (dX, dY). 

\subsection{Analysis results}
The primary focus of the 2022 VGOS-R\&D sessions was to provide proof that it is possible to significantly increase the number of scans per hour in VGOS sessions. 
Still, it is possible to also look at the performance of these sessions regarding geodetic parameter estimation and conduct comparisons with VGOS-OPS sessions. 
However, it has to be highlighted that the 2022 VGOS-R\&D series only includes six sessions. 
Thus, it is not possible to conduct a meaningful repeatability analysis and the subsequent comparisons need to be interpreted carefully. 

\begin{figure}[htb]
    \centering
    \includegraphics[width=\textwidth]{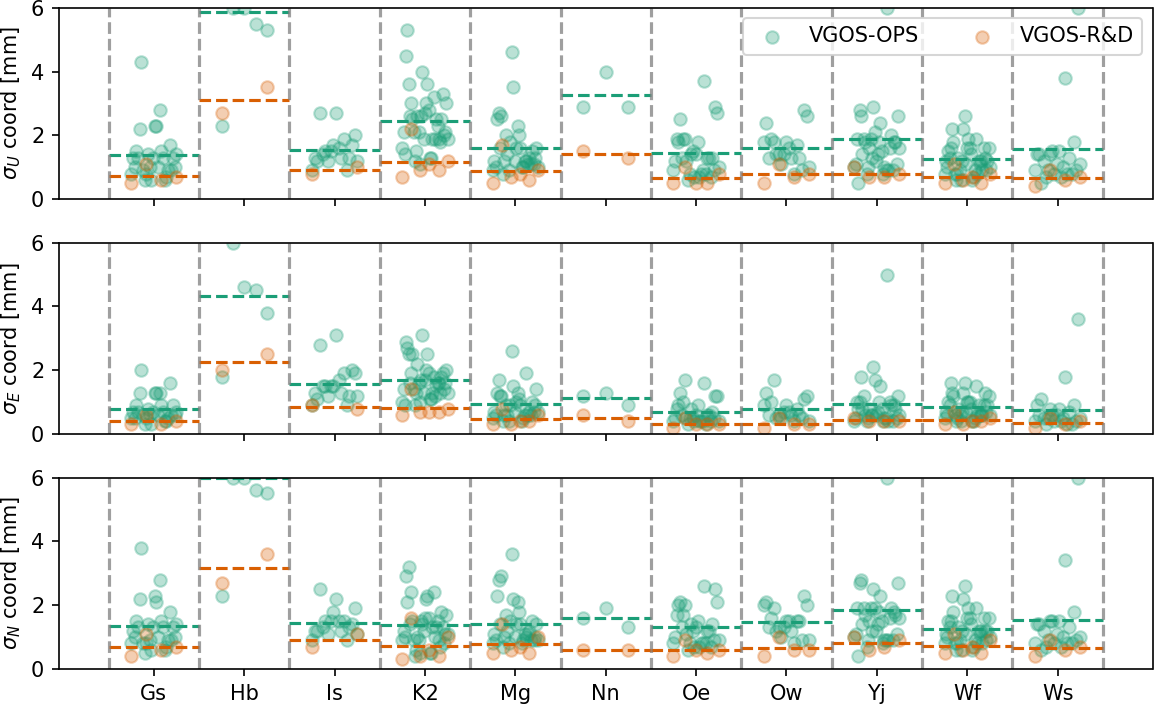}
    \caption{ Formal errors ($\sigma$) of station coordinates in up (top), east (mid), and north (bottom). 
    The markers represent individual sessions while the dashed lines depict the average. 
    VGOS-R\&D results are depicted in orange while VGOS-OPS results are depicted in turquoise. 
    Values larger than \SI{6}{mm} are displayed at \SI{6}{mm}.}
    \label{fig:coord}
\end{figure}

Figure~\ref{fig:coord} depicts the formal errors $\sigma$ of the estimated station coordinates based on the VieVS analysis. 
The average reduction of the formal errors per station is \SI{50}{\%} in all components.
Based on the results obtained from the official IVS analysis reports, the improvements are around \SI{40}{\%}.
Thus, both analysis approaches confirm a reduction of formal errors, which can be expected based on the greatly increased number of scans and thus observations available in the analysis. 
The improvement based on the VieVS results corresponds with the expected improvement from the Monte-Carlo simulations, while the improvement is smaller based on the IVS analysis report solution. 
The slight differences might be explained by the individual analysis settings used in the simulations compared to the analysis runs stemming from the different software packages that were used, or from imperfections in the simulations.

\begin{figure}
    \centering
    \includegraphics[width=\textwidth]{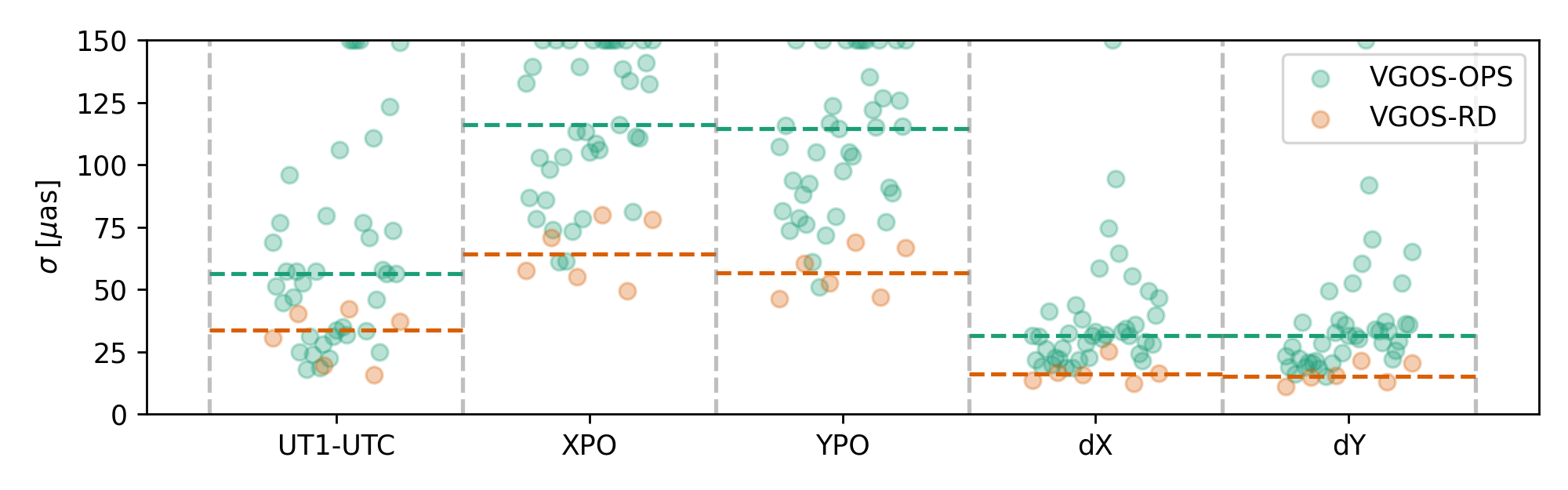}
    \caption{Formal errors ($\sigma$) of EOP. 
    The markers represent individual sessions while the dashed lines depict the median value. 
    VGOS-R\&D results are depicted in orange while VGOS-OPS results are depicted in turquoise. 
    Values larger than \SI{150}{\mu as} are displayed at \SI{150}{\mu as}.}
    \label{fig:eop}
\end{figure}

Figure~\ref{fig:eop} depicts the formal errors of the estimated EOP based on the VieVS analysis. 
In contrast to the previous sections, the median value is highlighted instead of the average one since there are a few significant outliers present. 
By comparing the median formal error per EOP, an average reduction of \SI{50}{\%} can be seen,  \SI{40}{\%} for UT1-UTC,  \SI{50}{\%} for polar motion, and  \SI{50}{\%} for the nutation parameters, which is in good agreement with the expectations based on the Monte-Carlo simulations. 
The improvement in terms of the formal errors of the EOP is in good agreement with the improvement in terms of the formal errors of the station coordinates discussed earlier, and also w.r.t. the expectations from the Monte-Carlo simulations. 

However, it has to be highlighted that the analyzed VGOS network is not well-suited for estimating EOP due to the lack of southern-hemisphere stations and consequently long north-south baselines. 
This fact might also explain the occurrence of some of the outlier values in Figure~\ref{fig:eop}. 

In this work, we omit a detailed analysis of the repeatability error due to the small sample size of the VGOS-R\&D sessions. 
Based on the investigations outlined in this work, we recommend applying the derived observation strategies in operational VGOS sessions to provide a larger sample size for statistically significant repeatability analysis. 

% ###################
% ### CONCLUSIONS ###
% ###################
\section{Conclusions} \label{sec:conclusion}
During 2022, the IVS provided resources for six dedicated 24-hour VGOS-R\&D sessions. 
These sessions aimed to greatly increase the number of observations and scans.
Compared to operational VGOS sessions, stations in VGOS-R\&D recorded 2.3 times more scans leading to 2.6 times more observations. 
This was achieved by establishing an SNR-based scheduling approach with shorter observation times of around 10 seconds and reduced overhead times. 

The SNR-based scheduling approach was tested based on two approaches. (1) based on inter-/extrapolation of existing S/X frequency source flux density models provided in the sked catalogs and (2) based on newly derived source flux density models for VGOS frequencies. 
In theory, we expected that a dedicated source flux density catalog at VGOS frequencies might be superior compared to inter-/extrapolating S/X models. 
However, both approaches resulted in a high observation success rate of \SIrange{90}{95}{\%}, assuming that no technical errors occur. 
The high success rate might be explained by the significant error margins included in the calculation of the required observation time. 

Comparisons of the theoretical predicted SNRs and the observed (reconstructed) SNRs revealed that the underlying models tend to be too optimistic, especially in band D. 
However, the existing error margins compensate for the mismodelling. 
By improving the underlying models, the current error margin in the calculation of the required observation time could be reduced to allow for even more aggressive scheduling strategies with shorter observation times. 
In any case, both approaches resulted in a significantly higher number of scans per hour compared to the current state-of-the-art strategy utilized in VGOS-OPS, where each scan is observed for 30 seconds straight. 

Despite the greatly increased number of scans per station, the reduced observation time led to an average reduction of recorded data per station of \SI{20}{\%}.
This reduction should lead to decreased data transfer and processing times, which represent the current bottleneck for advancing VGOS. 

Monte-Carlo simulations revealed that based on the updated scheduling strategy, the precision of the estimated station coordinates is expected to improve by \SI{50}{\%} compared to operational VGOS sessions. 
Similar improvement is expected for EOP estimates. 
Analysis of the 2022 VGOS sessions confirmed this assumption with \SI{50}{\%} lower formal errors in station coordinates based on analysis executed using VieVS, and \SI{40}{\%} lower formal errors based on the IVS analysis reports. 
The formal errors of EOP estimates were also reduced by \SIrange{40}{50}{\%}. 

\section{Outlook} \label{sec:outlook}
From an operational perspective, the SNR-based scheduling implemented in VGOS-R\&D, as well as the reduced overhead times, can be readily transferred to VGOS-OPS sessions. 
By recording more sessions with the increased number of scans per hour, a more sophisticated repeatability analysis can be executed to evaluate the impact on the accuracy of the estimated geodetic parameters to confirm the hypothesis outlined in the VGOS design documents. 

Utilizing a catalog with flux density models at VGOS frequencies would require operational updating, similar to the S/X catalog, due to the time-dependent nature of source brightness. 
Similarly, the station SEFD models need to be updated and monitored for VGOS frequencies and it must be ensured that these models are supported in the subsequent software packages. 

Finally, research by \citet{Anderson2018}, \citet{Xu2021a}, and \citet{Xu2021b} indicates that source structure effects also contribute significantly to the VGOS error budget. 
Work is underway to develop processes for imaging and modeling source morphology and for creating corrections for source structure and models that can be used for simulations. 
This necessitates consideration during the generation of VGOS schedules as discussed in \citet{Schartner2023}. 

\section*{Statements and Declarations}
\subsection*{Availability of data and materials}
The IVS-related datasets generated and/or analyzed during the current study are available in the IVS data centers and can be accessed via the corresponding session listed in  \url{https://ivscc.gsfc.nasa.gov/sessions/2022/}. 
The EOP dataset VIE2022 analyzed in this paper is available at \url{https://doi.org/10.48436/0gmbv-arv60}.
The source flux density catalog at VGOS frequencies developed for the VGOS-R\&D sessions is attached as supplementary material and in Appendix B. 

\subsection*{Competing interests}
The authors declare that they have no competing interests.

\subsection*{Funding}
The work by MIT Haystack Observatory was supported under NASA contract Awarding Agency.
\ \\
Open access funding is provided by the Swiss Federal Institute of Technology Zurich.

\subsection*{Authors' contributions}
MS and BP designed the concepts of the VGOS-R\&D sessions with support from MX. 
MS generated the observing plans.
MT correlated the sessions. 
MT, DH, and JB performed the fringe-fitting and post-correlation-processing.
DM performed data quality checks with support from DH, MT, and JB. 
MS performed the simulations. 
HK performed the geodetic analysis using VieVS. 
BP developed monitoring software to analyze the performance of stations and the brightness of sources. 
MS wrote the majority of the manuscript and summarized, compared, and visualized the results. 
All authors read and contributed to the manuscript. 

\subsection*{Acknowledgments}
This research has made use of VGOS data files provided by the International VLBI Service for Geodesy and Astrometry (IVS) data archives, dated July 2023.

\bibliography{sn-bibliography}% common bib file

\newpage

\section*{Appendix A: Error margins in VGOS observations}
It is important to note that due to imperfections in the source flux density and station $SEFD$ measurements and models, as well as since predictions of these metrics are required to perform the required calculation ahead of time, the calculation of the required observation time includes significant error margins to compensate for potential variations and mismodeling. 
Furthermore, it is to be noted that in VGOS sessions, all stations observe a scan for the same amount of time. 
Thus, only the least sensitive baseline on the least sensitive frequency band determines the required observation time. 
Consequently, only this baseline/band combination has a theoretical SNR level close to the target SNR, while all other baselines on all other bands typically have a higher SNR than the target, further increasing the error margin on these observations. 
As a side note, this also implies that the calculation of the observation duration and the underlying models are most important for small networks, in particular for single-baseline Intensives, which are beyond the focus of this work. 

The VGOS-OPS strategy of observing each scan for 30 seconds independent of the source brightness and station sensitivity can be seen as a further increase in the error margin for the VGOS observation time calculation due to the increased SNR. 
However, while the error margins help to ensure that planned observations can be used in the analysis, they also result in increased and potentially unnecessary observation duration. 
Thus, higher error margins lead to a lower number of scans per hour. 
Furthermore, the longer observation duration also leads to more recorded data that needs to be transferred and processed, which is currently the major bottleneck in operational VGOS VLBI, limiting the number of sessions that can be observed per month. 
The increased observation duration reduces the number of scans that can be executed within one session. 

In summary, the error margin can be grouped into three areas: (1) a general error margin to compensate for unforeseen minor technical issues at the telescopes or other unforeseen factors such as radio frequency interference, (2) one to reflect errors in source flux density models and to cover the uncertainty in the predictions, and (3) one to reflect errors in station sensitivity models and to cover natural variations in the predictions. 
By improved modeling, (2) and (3) can be reduced while some amount of error margin from (1) will need to remain, e.g. since it is common that stations have small technical problems that result in sensitivity losses of some tens of percent in some bands which need to be compensated. 
Since the distinction between these three groups is only theoretical, the choice of (1) determines the accuracy level required for the $SEFD$ and $F$ models. 
In simple terms, if an operator decides to account for \SI{50}{\%} total error margin, there is little need to spend significant resources to bring the accuracy of the $SEFD$ and $F$ models down to a few percent. 
The main objective is to ensure that the total error margin is reasonable to provide a good compromise between the observation success rate and the number of scans that can be observed. 
%In particular, within the VGOS-R\&D project, two approaches to provide source flux density models with sufficient accuracy were explored since station SEFD measurements are already executed regularly at many stations. 
\newpage
\begin{landscape}

\section*{Appendix B: Source flux density table}
Source flux density catalog at VGOS frequencies derived for the 2022 VGOS-R\&D sessions. 
The columns represent the projected baseline length in km up to which the listed flux density values correspond. 
Please note that this catalog was derived by analyzing VGOS observations up to 2022. 
Since source brightness is known to vary over time, there is no guarantee that the provided values are accurate enough for operational VLBI scheduling past 2022. 
\footnotesize
\begin{longtable}{llrrrrrrrrrrrrr}
\toprule
 & [km] & 1000 & 2000 & 3000 & 4000 & 5000 & 6000 & 7000 & 8000 & 9000 & 10000 & 11000 & 12000 & 13000 \\
Source & &  &  &  &  &  &  &  &  &  &  &  &  &  \\
\midrule
\endfirsthead
\toprule
 & [km] & 1000 & 2000 & 3000 & 4000 & 5000 & 6000 & 7000 & 8000 & 9000 & 10000 & 11000 & 12000 & 13000 \\
Source &  &  &  &  &  &  &  &  &  &  &  &  &  &  \\
\midrule
\endhead
\midrule
& & \multicolumn{9}{l}{source flux density up to projected baseline length} & \multicolumn{4}{r}{Continued on next page} \\
\midrule
\endfoot
\midrule
& & \multicolumn{9}{l}{source flux density up to projected baseline length} \\
\bottomrule
\endlastfoot
\multirow[t]{4}{*}{0003-066} & A & 2.24 & 2.06 & 1.92 & 1.84 & 1.80 & 1.76 & 1.70 & 1.61 & 1.49 & 1.35 & 1.22 & 1.08 & 0.95 \\
 & B & 3.36 & 3.19 & 3.01 & 2.82 & 2.62 & 2.41 & 2.19 & 1.95 & 1.67 & 1.37 & 1.08 & 0.78 & 0.48 \\
 & C & 3.62 & 3.43 & 3.21 & 2.96 & 2.71 & 2.45 & 2.19 & 1.91 & 1.62 & 1.30 & 0.99 & 0.68 & 0.36 \\
 & D & 3.42 & 3.18 & 2.89 & 2.58 & 2.25 & 1.92 & 1.61 & 1.33 & 1.07 & 0.84 & 0.61 & 0.37 & 0.14 \\
\midrule
\multirow[t]{4}{*}{0016+731} & A & 0.94 & 0.88 & 0.81 & 0.75 & 0.68 & 0.61 & 0.54 & 0.46 & 0.38 & 0.30 & 0.22 & 0.15 & 0.07 \\
 & B & 1.07 & 1.02 & 0.95 & 0.88 & 0.79 & 0.70 & 0.61 & 0.53 & 0.47 & 0.41 & 0.36 & 0.31 & 0.27 \\
 & C & 1.17 & 1.09 & 1.00 & 0.91 & 0.81 & 0.73 & 0.65 & 0.59 & 0.54 & 0.50 & 0.46 & 0.42 & 0.38 \\
 & D & 1.20 & 1.11 & 1.02 & 0.93 & 0.86 & 0.80 & 0.75 & 0.71 & 0.68 & 0.66 & 0.65 & 0.63 & 0.61 \\
\end{longtable}
\end{landscape}

\section*{Appendix C: Theoretical (predicted) SNR per session and baseline}

\begin{figure} [H]
    \centering
    \includegraphics[width=.7\textwidth]{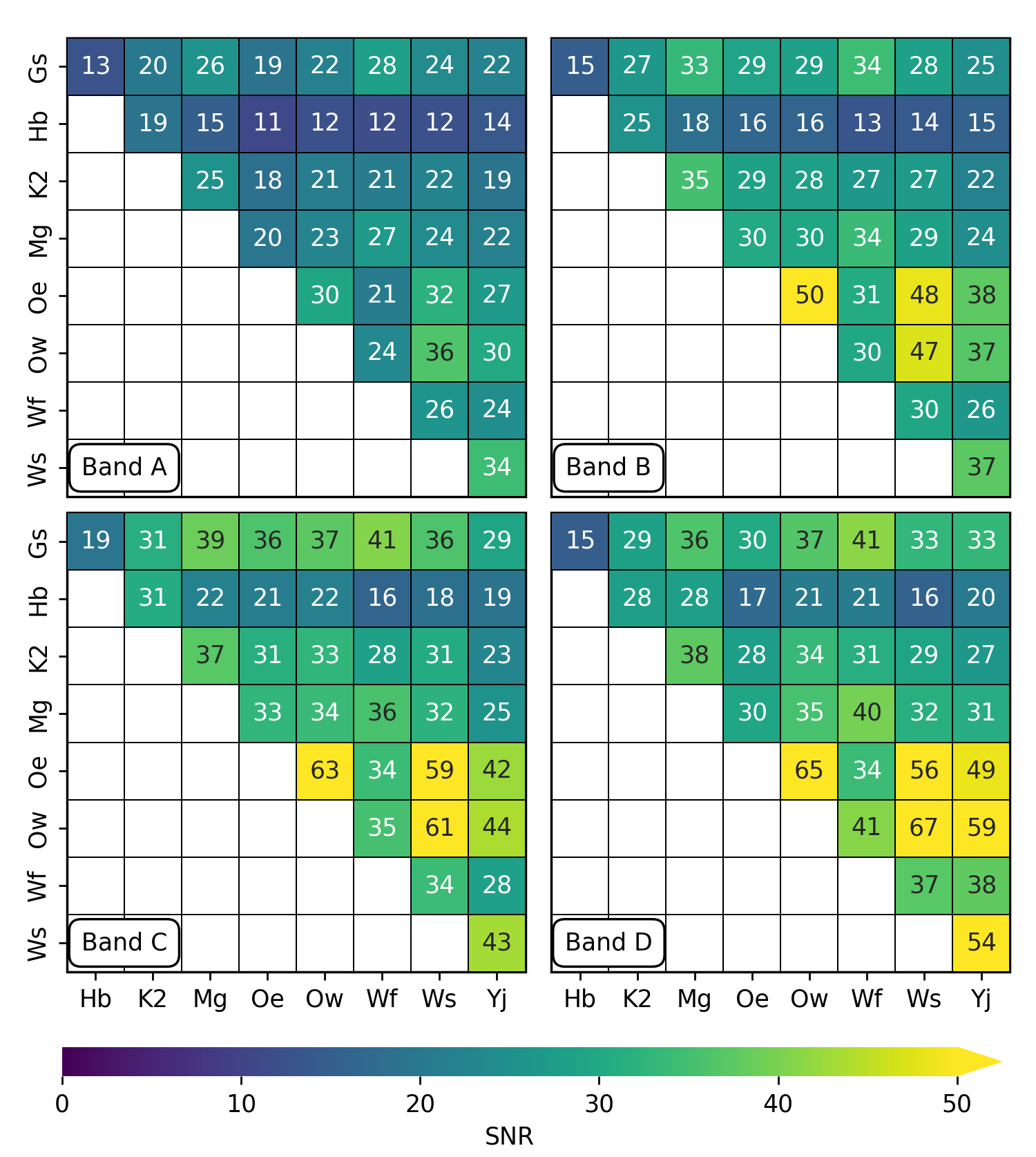}
    \caption{Average predicted SNR per baseline and band for session VR2201.}
%    \label{fig:snr_obs_vr2203}
\end{figure}

\begin{figure} [H]
    \centering
    \includegraphics[width=.7\textwidth]{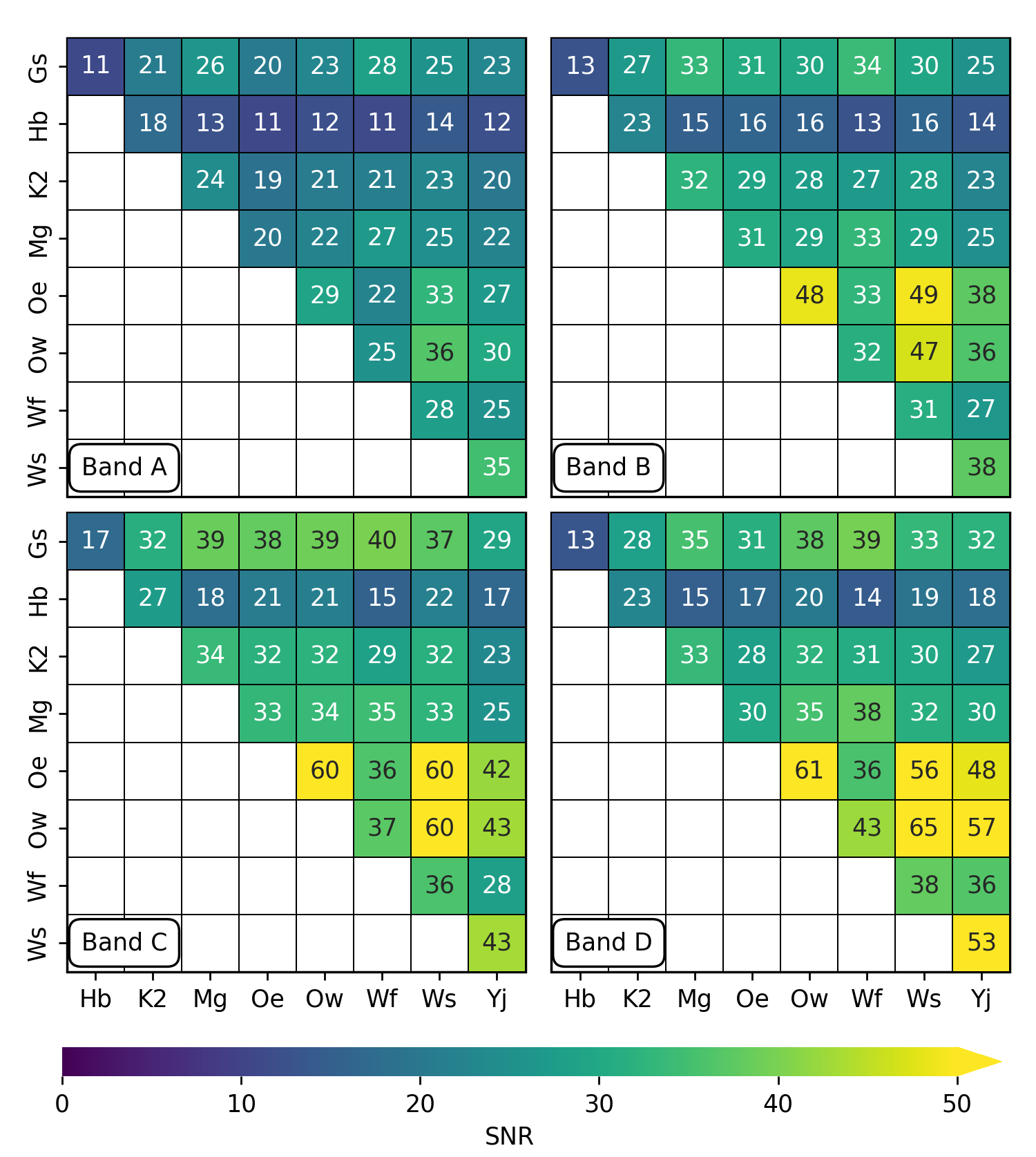}
    \caption{Average predicted SNR per baseline and band for session VR2202.}
%    \label{fig:snr_obs_vr2203}
\end{figure}

\begin{figure} [H]
    \centering
    \includegraphics[width=.7\textwidth]{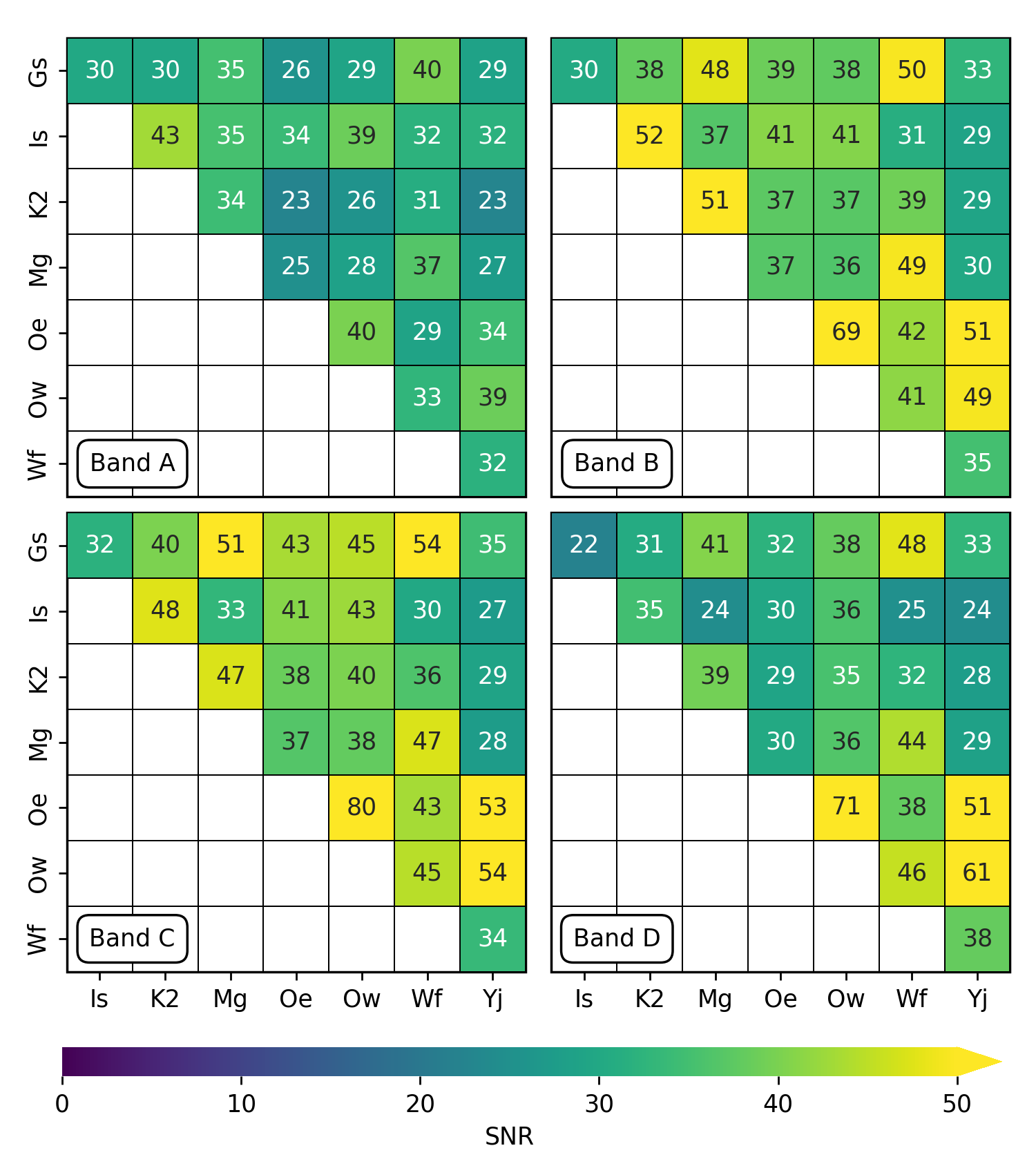}
    \caption{Average predicted SNR per baseline and band for session VR2203.}
%    \label{fig:snr_obs_vr2203}
\end{figure}

\begin{figure} [H]
    \centering
    \includegraphics[width=.7\textwidth]{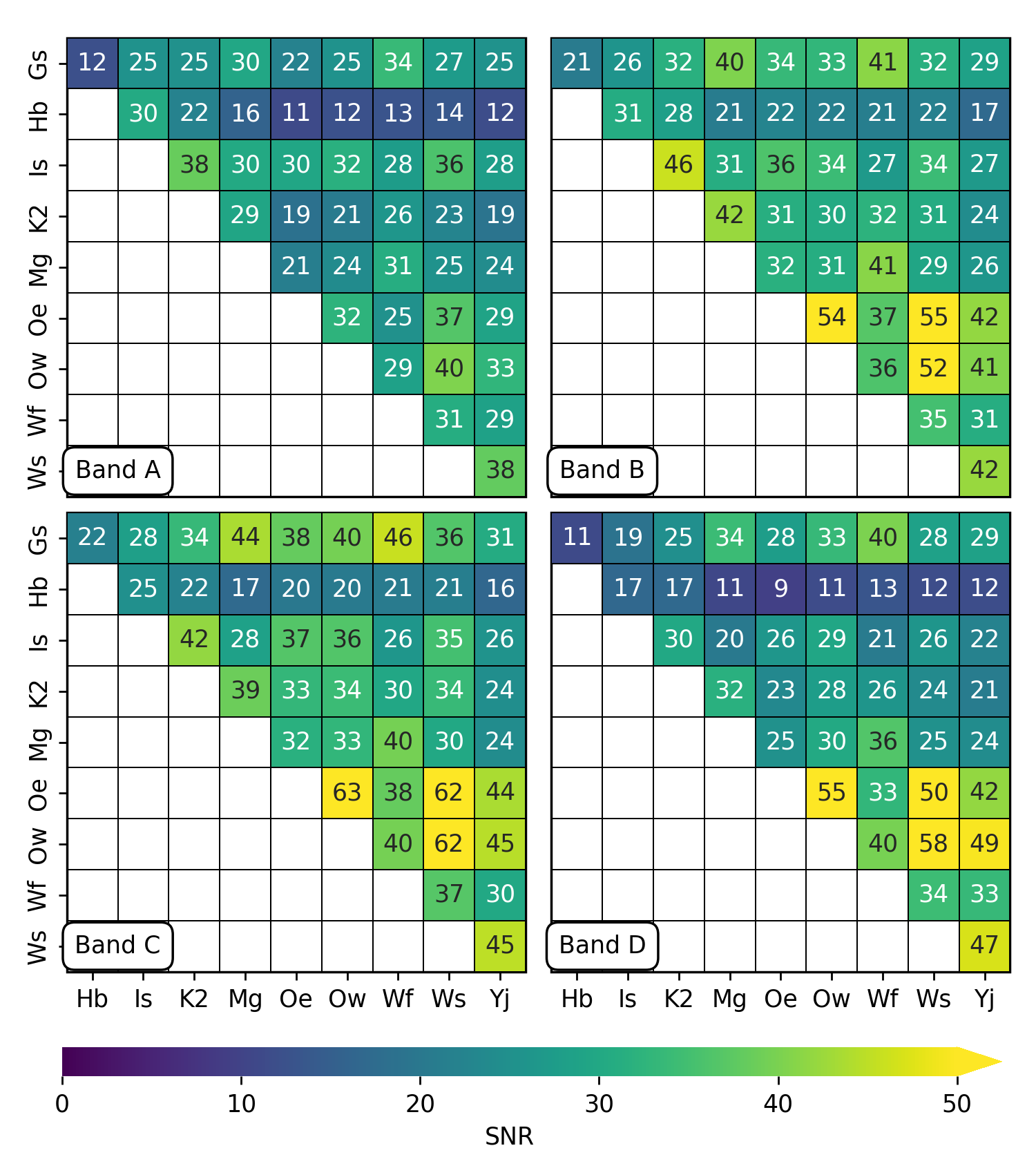}
    \caption{Average predicted SNR per baseline and band for session VR2204.}
%    \label{fig:snr_obs_vr2203}
\end{figure}

\begin{figure} [H]
    \centering
    \includegraphics[width=.7\textwidth]{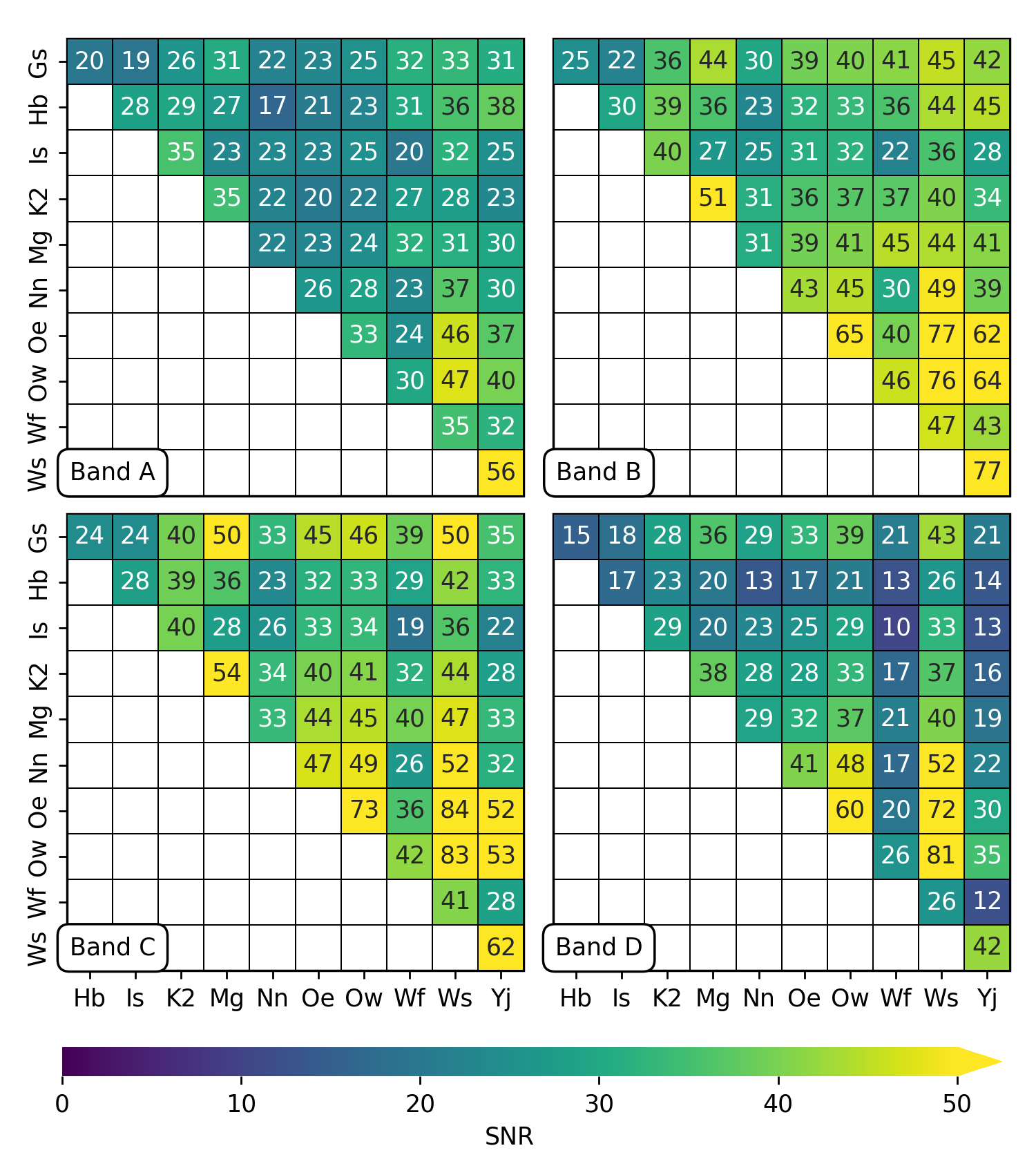}
    \caption{Average predicted SNR per baseline and band for session VR2205.}
%    \label{fig:snr_obs_vr2203}
\end{figure}

\begin{figure} [H]
    \centering
    \includegraphics[width=.7\textwidth]{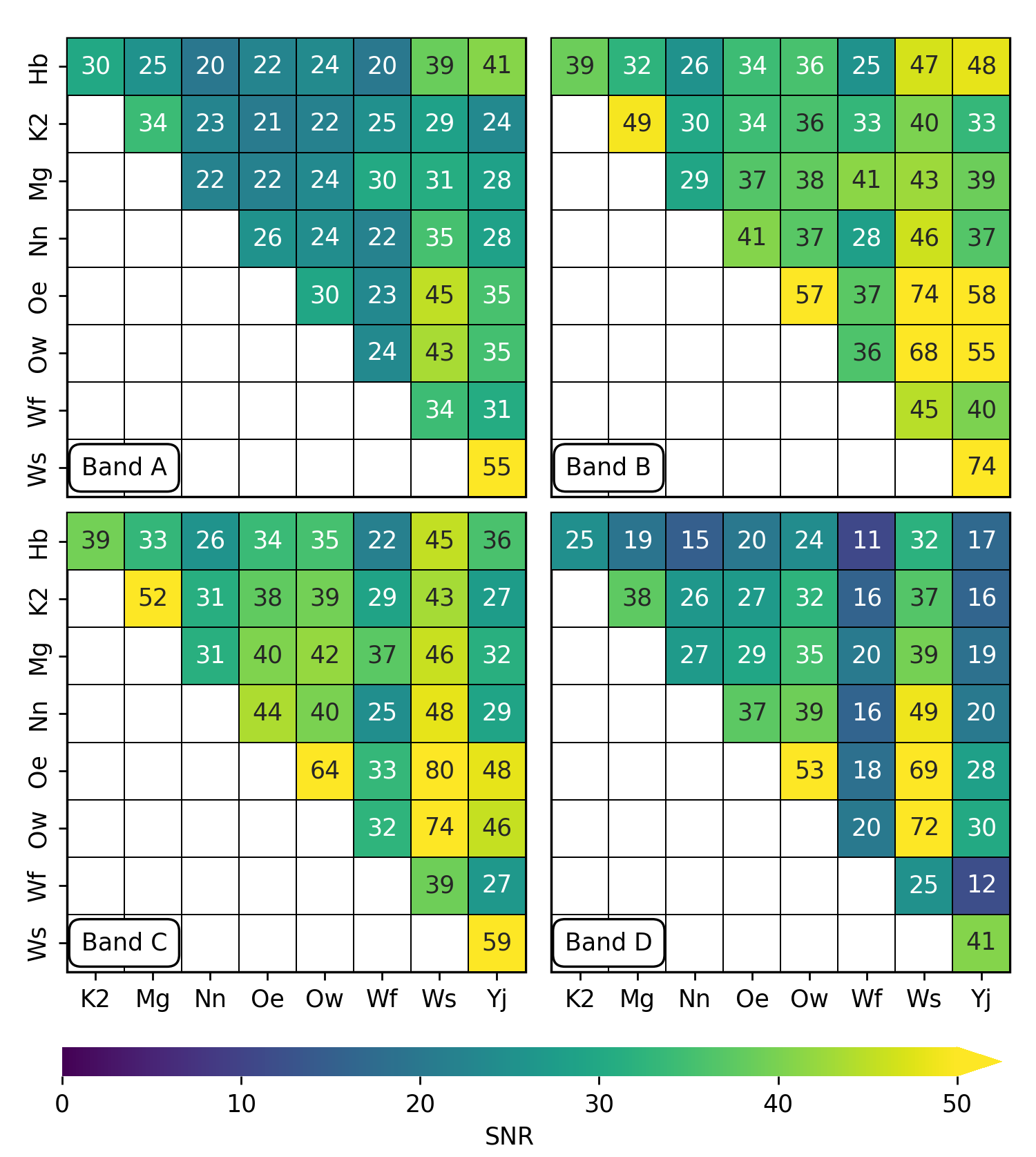}
    \caption{Average predicted SNR per baseline and band for session VR2206.}
%    \label{fig:snr_obs_vr2203}
\end{figure}

\newpage
\section*{Appendix D: Observed (reconstructed) SNR per session and baseline}
\begin{figure} [H]
    \centering
    \includegraphics[width=.7\textwidth]{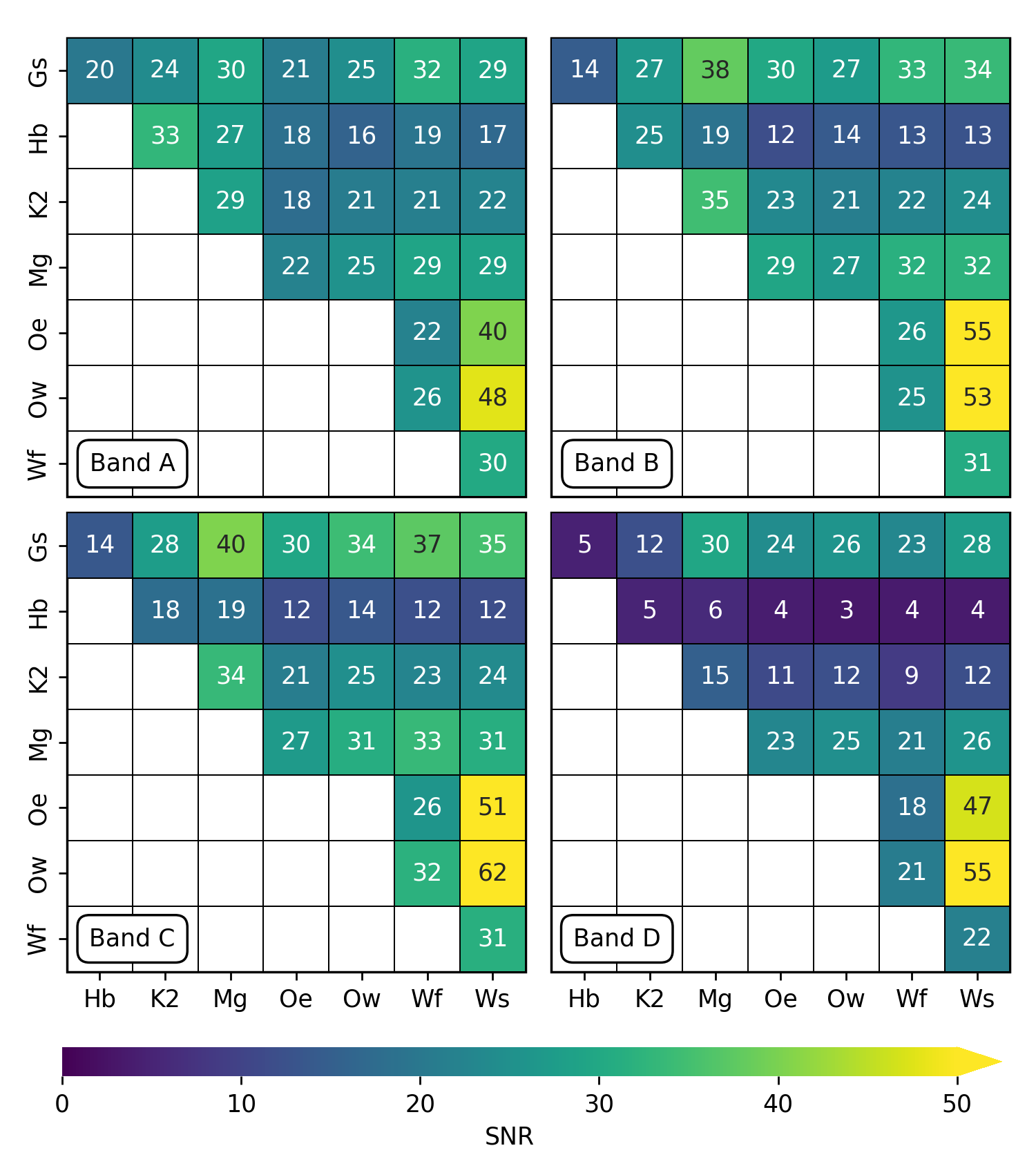}
    \caption{Average reconstructed SNR per baseline and band for session VR2201.}
%    \label{fig:snr_obs_vr2203}
\end{figure}

\begin{figure} [H]
    \centering
    \includegraphics[width=.7\textwidth]{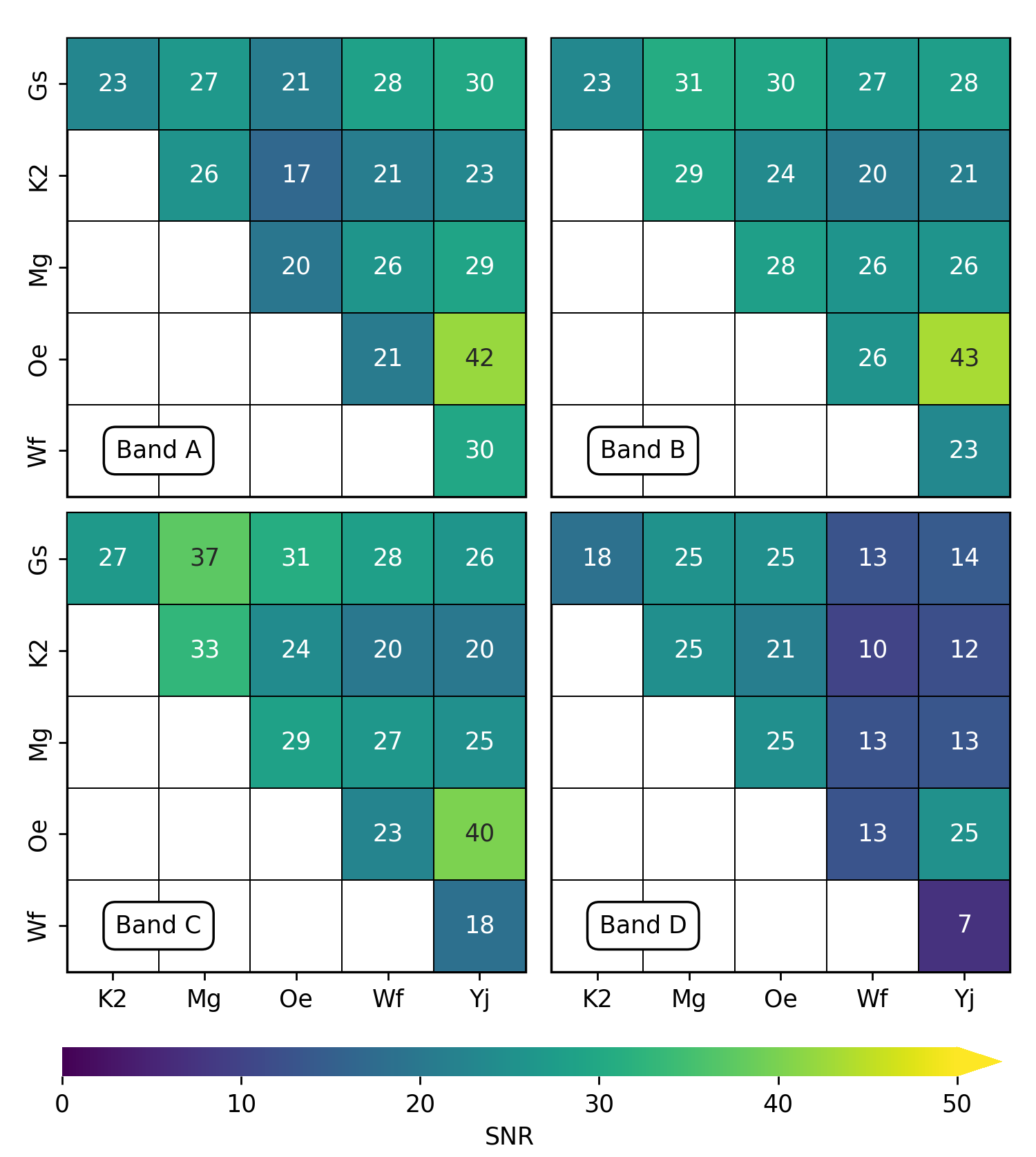}
    \caption{Average reconstructed SNR per baseline and band for session VR2202.}
%    \label{fig:snr_obs_vr2203}
\end{figure}

\begin{figure} [H]
    \centering
    \includegraphics[width=.7\textwidth]{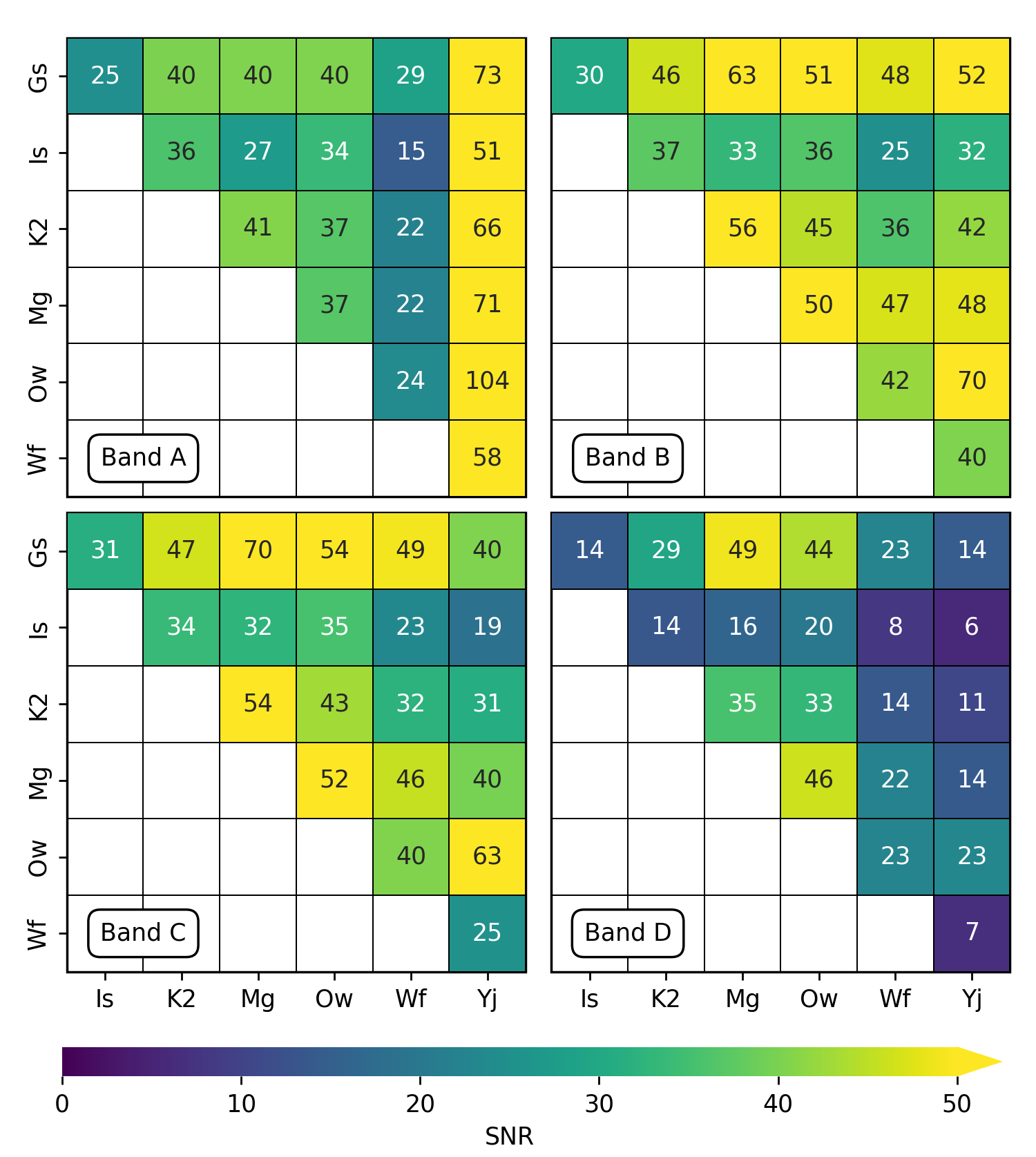}
    \caption{Average reconstructed SNR per baseline and band for session VR2203.}
%    \label{fig:snr_obs_vr2203}
\end{figure}

\begin{figure} [H]
    \centering
    \includegraphics[width=.7\textwidth]{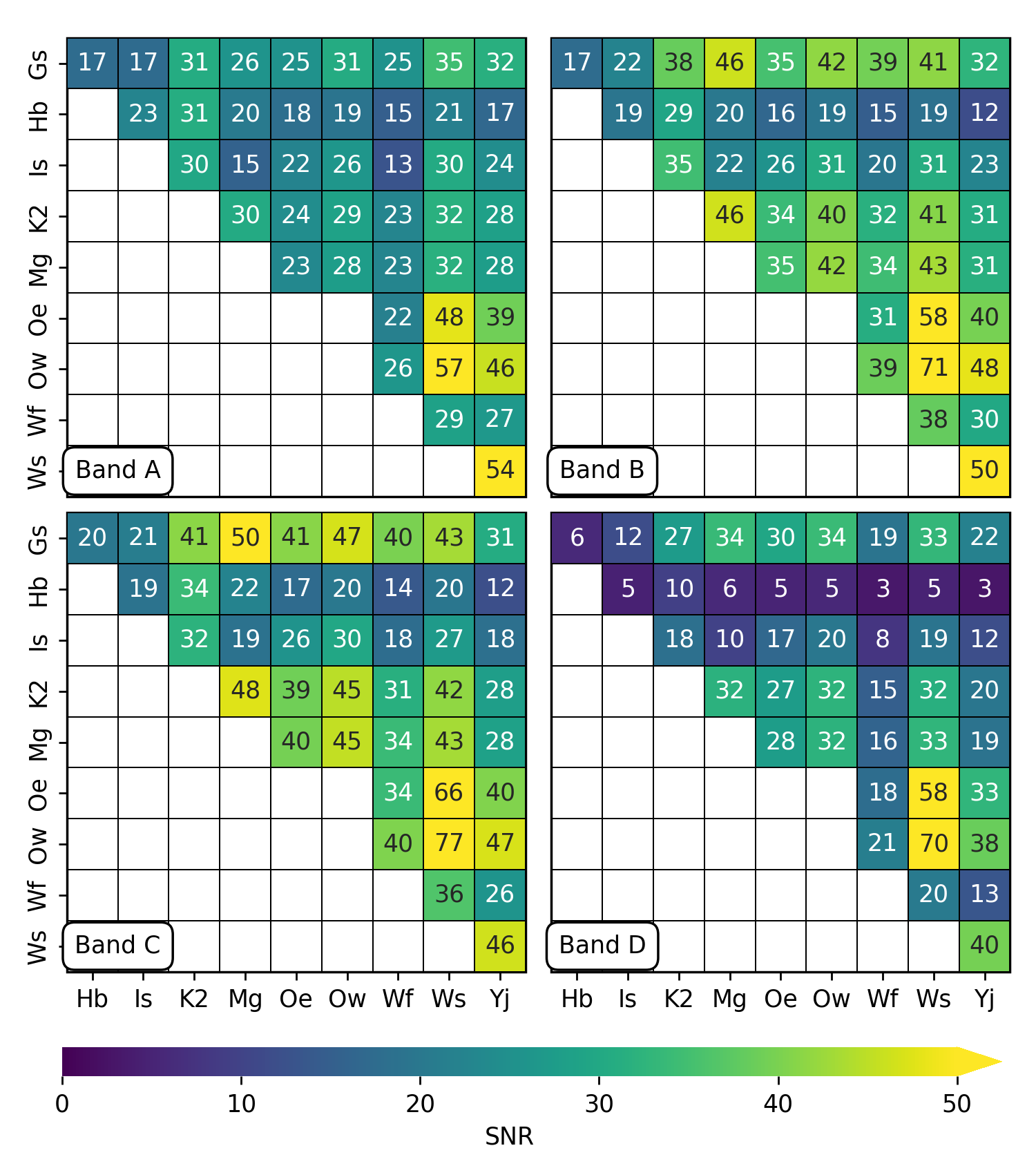}
    \caption{Average reconstructed SNR per baseline and band for session VR2204.}
%    \label{fig:snr_obs_vr2203}
\end{figure}

\begin{figure} [H]
    \centering
    \includegraphics[width=.7\textwidth]{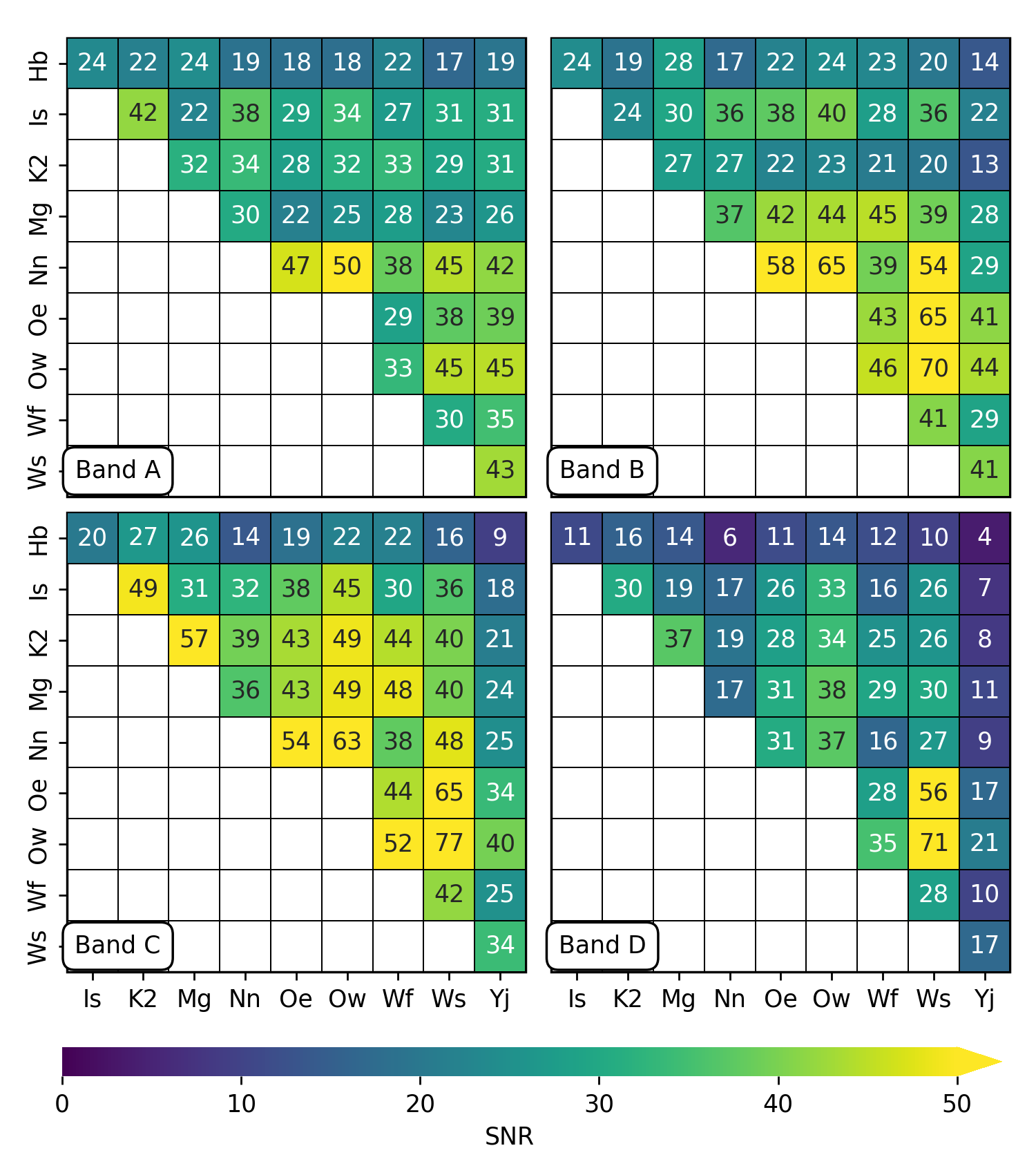}
    \caption{Average reconstructed SNR per baseline and band for session VR2205.}
%    \label{fig:snr_obs_vr2203}
\end{figure}

\begin{figure} [H]
    \centering
    \includegraphics[width=.7\textwidth]{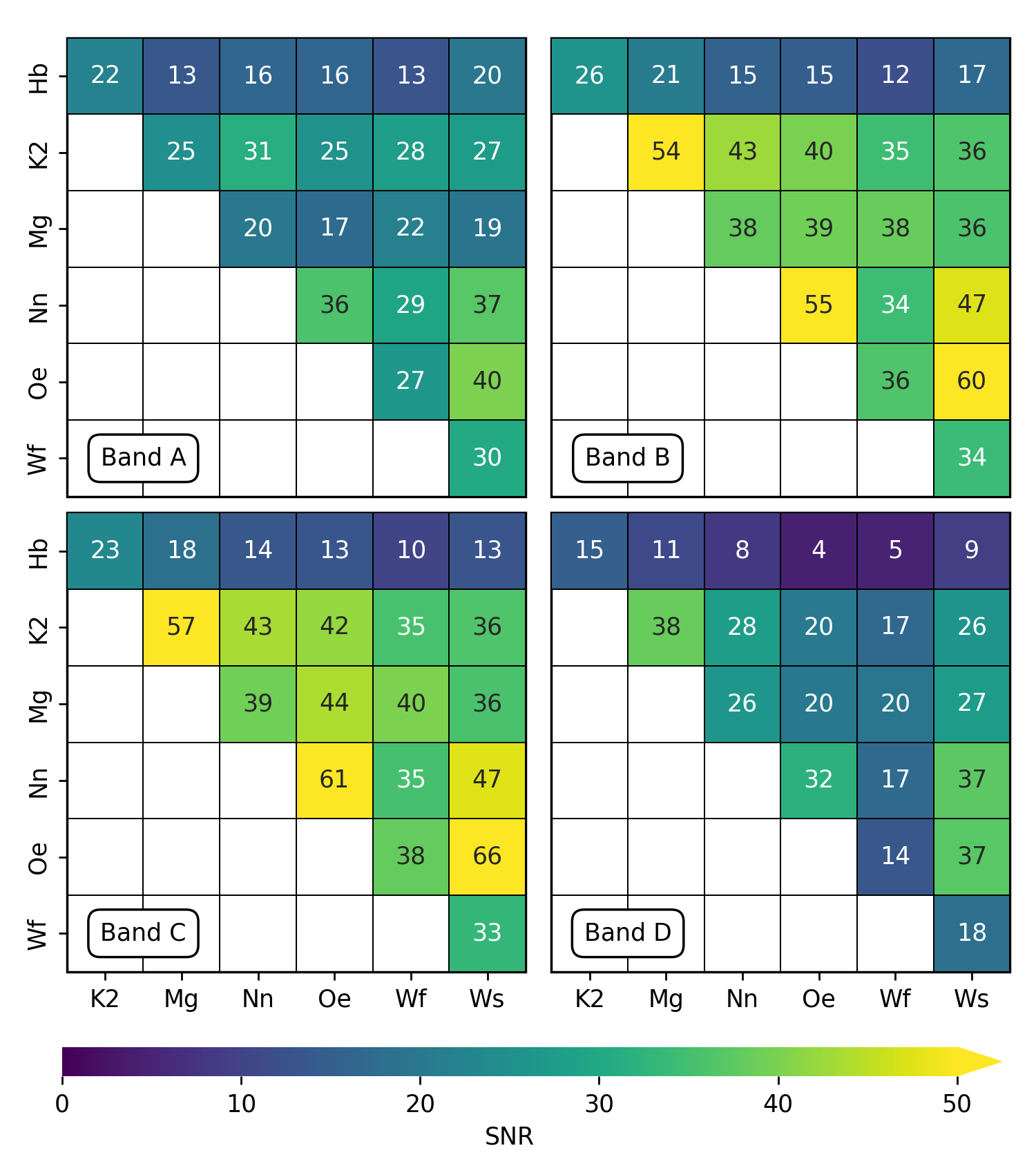}
    \caption{Average reconstructed SNR per baseline and band for session VR2206.}
%    \label{fig:snr_obs_vr2203}
\end{figure}

\end{document}